\documentstyle[twocolumn,prd,aps,eqsecnum,floats,epsfig]{revtex}
\tighten
\begin{document}
\title{Relativistic unitary description of $\pi\pi$ scattering}
\author{M.~A.~Pichowsky, A.~Szczepaniak, and J.~T.~Londergan}
\address{Department of Physics and Nuclear Theory Center, 
	Indiana University, Bloomington, Indiana 47405}
\maketitle
%------------------- Abstract ------------------------------------------%
\begin{abstract}
A unitary framework based on the Bakamjian-Thomas construction of
relativistic quantum mechanics is used to describe two-pion scattering
from threshold to 1400~MeV.  
The framework properly includes unitarity cuts for one-, two- and
three-hadron states and provides an excellent description of the available
data for $\pi\pi$ phase shifts and inelasticities.
The role and importance of three-hadron cuts are calculated and discussed. 
\end{abstract}
%
%------------------- Main Body ------------------------------------------
\def\k{{\bf k}}
\def\q{{\bf q}}
\def\p{{\bf p}}
\def\s{{\bf s}}
\def\P{{\bf P}}
\def\Q{{\bf Q}}
\def\K{{\bf K}}
\def\E{{\cal E}}
\def\M{{\cal M}}
\def\del3#1#2{(2\pi)^3\delta({#1}\!-\!{#2})}
\def\dek#1#2{\delta_{{#1}{#2}}}
\def\sfrac#1#2{{\textstyle{{#1}\over {#2}}}}  % small fraction style!
\def\Re{\hbox{Re}\,}
\def\Im{\hbox{Im}\,}
\def\Arg{\hbox{Arg}\,}
\def\bra#1{\langle{#1}|}
\def\ket#1{|{#1}\rangle}
\def\bracket#1#2{\langle{#1}|{#2}\rangle}
\newcommand{\beq}{\begin{equation}}
\newcommand{\eeq}{\end{equation}}
\newcommand{\bea}{\begin{eqnarray}}
\newcommand{\eea}{\end{eqnarray}}
\newcommand{\rf}[1]{(\ref{#1})}
\def\al{\alpha}
\def\be{\beta}
\def\2p{ (2\pi)^3}
\def\ie{{\it i.e.}}
\def\eg{{\it e.g. }}
\newcommand{\Eq}[1]{Eq.~(\ref{#1})}
\newcommand{\Eqs}[1]{Eqs.~(\ref{#1})}
\newcommand{\Ref}[1]{Ref.~\cite{#1}}
\newcommand{\Refs}[1]{Ref.~\cite{#1}}
\newcommand{\Sec}[1]{Sec.~\ref{#1}}
\newcommand{\Secs}[1]{Secs.~\ref{#1}}
\newcommand{\Fig}[1]{Fig.~{\ref{#1}}}
\newcommand{\Figs}[1]{Figs.~{\ref{#1}}}
\renewcommand{\+}{\!+\!} 
\renewcommand{\-}{\!-\!}
\newcommand{\tpc}{(2\pi)^3}
%........................................................
%......................................................................
\section{Introduction}
A nonperturbative framework capable of describing the relativistic,
coupled-channel scattering of hadrons is presented. 
The approach is based on a relativistic Hamiltonian formulation 
with model interactions introduced into the mass operator,
and with few-body states implemented in a way that maintains the unitarity
of the theory.
The elementary degrees of freedom in the framework are finite-sized 
hadrons which provide a natural ultraviolet regularization, ensuring that  
the scattering amplitudes are finite.

The Hilbert space is truncated to include only one-, two- and three-body
states.
A central and novel feature of this framework is the explicit inclusion of
both real and imaginary parts of scattering amplitudes arising from the
opening of three-body channels. 
The proper handling of three-body unitarity cuts is crucial to gaining
a deeper understanding of several well-known scattering systems; a good
example is $\pi N$ scattering in the $P_{11}$ channel, which
exhibits a significant inelasticity arising from the intermediate
three-body $\pi\pi N$ state~\cite{Lee86}.

It presents a formidable challenge to develop a general, 
relativistic scattering framework to describe the final-state 
interactions between hadrons, that includes the
effects of three-body unitarity cuts. 
Nonetheless, a practical framework which can treat hadron reactions beyond
the lowest-order valence quark picture is clearly desirable.
For example, the systematic analysis of hadron reactions in the baryon
resonance region currently being conducted in Hall B at the Thomas
Jefferson National Accelerator Facility (TJNAF) requires that 
such a framework be used to extract information about baryon
resonances in this highly complex dynamical region.  
The framework developed herein is an attempt to construct a useful,
relativistic framework capable of describing the nonperturbative,
low-momentum transfer final-state interactions between hadrons in a
unitary manner. 

For the first application of the framework developed here, a simple model
for $\pi\pi$ scattering is introduced and used to described the $S$ and
$P$ partial waves for energies ranging from the two-pion threshold up to
1400~MeV.   
This system provides an excellent test for the framework.
A relativistic treatment is quite important when dealing with particles as
light as pions and the interplay between strong dynamics and chiral
symmetry makes this system quite interesting.
The $\pi\pi$ system is somewhat simpler than others, in that its study
requires only a minimal complication from the proper implementation of
relativistic spins, since both of the two-body states ($\pi\pi$ and $K\bar{K}$)
involved are comprised of spin-0 particles.  
Another attractive aspect of applying the framework to $\pi\pi$
scattering is the relative wealth of experimental data for the isoscalar,
$S$-wave channel.   

One drawback with using $\pi\pi$ scattering as a touchstone
may be that inelasticities due to {\em open} states of three or more
particles, do not appear to be significant for this process;  
that is, the $\rho\pi\pi$ and $\pi\pi\pi\pi$ thresholds seem to have
little impact on the $S$- and $P$-wave observables. 
The effects of the opening of these three- and four-body channels 
seem to be overwhelmed by the opening of the two-body $K\bar{K}$ channel. 
Nonetheless, several important aspects of the framework can be explored in
an application to $\pi\pi$ scattering.  

The isoscalar-scalar ($I=0$, $J^{PC}=0^{++}$) channel of $\pi\pi$
scattering has been a subject of numerous and extensive studies.  
The study of meson scattering in this low-energy region may be an ideal
test of our understanding of the interplay between bound states in QCD and
chiral dynamics.  
The region near $E=$ 1000~MeV is perhaps most interesting, as it is 
dominated by the mixing between $\pi\pi$ and $K\bar{K}$ channels and the
isoscalar-scalar $f_0(980)$ meson resonance.
The nature of the $f_0(980)$ resonance, and the question of whether it is
comprised of valence quarks or arises purely from meson scattering
dynamics, has been addressed by many 
authors~\cite{f0980,f0980:1,f0980:2,f0980:3,f0980:4,f0980:5,f0980:6}.
Above this energy region, three additional scalar meson resonances have
been well established.  
These are referred to as the $f_0(1370)$, $f_0(1500)$ and the
$f_0(1710)$.   It is still unclear which of these should be considered 
as quark-antiquark bound states, glueballs or possibly resonances arising 
from dynamical effects of final-state
interactions~\cite{PDG,1gev:latt,1gev:1,1gev:2,1gev:3}.  
Thus, the isoscalar-scalar channel remains a source of great interest and
mystery for meson phenomenology.

Although there have been previous studies which employ a framework similar
to the one developed here, there are some important differences.
Most studies of meson scattering dynamics are based on potential models  
(as is the framework developed here.)  
However, most other approaches typically include one- and two-particle
channels only; that is, they include $s$-channel states and 
several two-particle channels, such as $\pi\pi$, $K\bar K$,
$\sigma\sigma$, etc.  
They either neglect the possibility of open three-particle channels
altogether or only partially implement them.
For example, in the model developed by the Julich group\cite{Speth95}, the 
interaction potentials between two-particle channels, such as
$\pi\pi$-$\pi\pi$ or $\pi\pi$-$K \bar{K}$ interactions are obtained using
an instantaneous approximation of a meson-exchange model. 
Such instantaneous approximations generally do not account for absorptive
effects due to the opening of three-body channels. 
Still, there is no question that the Julich model quite successfully
describes the phase shifts and inelasticities of $\pi\pi$ scattering 
for $S$, $P$ and $D$ waves.
Alternatively, the Krakow group~\cite{Kaminski94} has developed a
separable-potential model for $\pi\pi$ scattering.  
In their calculation, few-body dynamical effects are incorporated by
including additional, {\em effective} two-body channels, such as a
$\sigma\sigma$ channel~\cite{ss}.  
Their model also obtains excellent results for the $\pi\pi$ phase shifts 
and inelasticities.

The outline of this article is as follows.  
In \Sec{Sec:Formalism},  the relativistic scattering formalism employed
herein is briefly discussed, beginning with a short proof of the
covariance of observables calculated within this framework.
Then, the integral equations that relate the one-, two- and
three-body scattering $T$-matrices are provided.

In \Sec{Sec:PiPi}, the framework is applied to a study of
$\pi\pi$ scattering.  
The particle states that are included in the model are discussed, along
with the necessary dynamical model parameters.
The interactions employed in this study arise from the meson exchanges
which couple states of various numbers of particles to each other.
In our framework, these interactions arise from one-, two- and three-meson
intermediate states which may exhibit production thresholds, resulting in
absorptive contributions to the kernels and self-energies appearing in the
Lippmann-Schwinger equations.
A simple fitting procedure is shown to provide excellent agreement with
data for $\pi\pi$ scattering phase shifts and inelasticities.
Details of the relevant model dynamics that produce the various features
observed in the resulting phase shifts and inelasticities are discussed.
Then, it is shown that the numerical methods employed herein are
sufficient to maintain the unitarity of the framework to better than one
part in a million. 
Finally, in \Sec{Sec:Summary}, the article is summarized and plans for
future studies are presented.

%------------------------------------------------------------------------
\section{Relativistic Quantum Mechanical Framework} 
\label{Sec:Formalism}
In this section, a relativistic Hamiltonian framework that provides a
covariant unitary approach to the study of multichannel  scattering is
described.  Lorentz symmetry is maintained by 
identifying the interactions with the mass operator 
(that is, the Hamiltonian in the overall center-of-momentum frame).  
It is shown in \Ref{Bakamjian} that the complete set of Poincare
generators can be constructed in a simple way that separates the internal
dynamics from the center-of-momentum (CM) motion. 
In \Sec{Sec:RelCo} a proof of the covariance of this approach is provided.
Furthermore,  Betz and Coester \cite{Coester80} show that such a framework
can satisfy cluster separability.
All of these features are desirable for the study of hadron scattering.

In \Sec{Sec:RelCo}, the Lorentz covariance of the framework is
demonstrated and the fully-interacting mass operator $\M$ is constructed.
It is shown that the framework leads to Lorentz-invariant on-shell
T-matrix elements $T(E,{\bf P})$ for colliding particles with 
total momentum ${\bf P}$ and energy $E$; 
that is, one finds $T(E,{\bf P})=T(\sqrt{s})$ where 
$\sqrt{s}=\sqrt{E^2-\P^2}$ is the invariant mass of the system.

The Poincare generators act on a Hilbert space which, in general, contains
an infinite number of states.
The Hilbert space is truncated to include only those states 
essential to describe the scattering system of interest within a
particular energy range.   
Here, only one-, two- and three-particle states are maintained.
Following this truncation, the operator form of the Lippmann-Schwinger 
equation can be written as a set of coupled integral equations.  
The input that determines the dynamics is given in terms of
the matrix elements of a model potential $V$.  
Once these are provided, the full scattering problem is solved in a
straightforward manner.   

%---------------------------------------------------------------------
\subsection{Relativistic covariance}
\label{Sec:RelCo}

A simple realization of the Poincare algebra for an interacting system
of a finite number of constituents is given by the Bakamjian-Thomas
construction~\cite{Bakamjian}. 
This approach has the advantage of providing a Lorentz-covariant
generalization for a large class of noncovariant microscopic models,  
such as the constituent quark model.
In principle, a noncovariant microscopic model could be used to obtain
matrix elements of the underlying elementary hadronic potentials $V$.
In this case, one might consider this framework as a means to extend the
original noncovariant model dynamics, allowing for a Lorentz-covariant
treatment of scattering phenomena.    

The explicit construction of the Poincare algebra proceeds as
follows.  Starting from a system of {\em noninteracting} particles,
described by their coordinates ${\bf x}_a$, momenta ${\bf p}_a$, 
spins ${\bf s}_a$, and masses $m_a$, the Poincare generators are
\begin{eqnarray}
H &=& \sum_a \E(m_a,\p_a) = \sum_a\sqrt{m_a^2 + \p_a^2}, 
	 \nonumber \\ 
{\bf P} &=& \sum_a {\bf p}_a,  \nonumber \\
{\bf J} &=& \sum_a {\bf x}_a\times {\bf p}_a + {\bf s}_a, \nonumber \\ 
{\bf K} &=& \sum_a {1\over 2}\{  {\bf x}_a, \E(m_a,\p_a) \} 
	- { {{\bf s}_a \times {\bf p}_a }
	\over {\E(m_a,\p_a) + m_a}}.
\label{Def:Edef} 
\end{eqnarray}
Here, $H$ and ${\bf P}$ are the total free energy and linear 
momentum of the system, ${\bf J}$ and ${\bf K}$ are the total angular
momentum and boost operators, respectively. 
The relative coordinates ${\bf r}_a$, relative momenta ${\bf k}_a$,
center-of-momentum (CM) spins ${\bf s}'_a$,  and the CM coordinates
${\bf R}_{cm}$, total momentum ${\bf P}_{cm}$, and total spin 
${\bf S}_{cm}$,  are introduced via the Gartenhaus-Schwartz transformation
which allows a separation of the {\em internal} dynamics and CM motion.
In terms of these new variables, the Poincare generators are given by 
\begin{eqnarray}
H &=& \sqrt{{\bf P}^2 + {\cal M}^2(\k_1,\k_2,\ldots)}, \nonumber \\ 
{\bf P} &=& {\bf P}_{\rm cm},  \nonumber \\
{\bf J} &=& {\bf R}_{\rm cm}\times {\bf P}_{\rm cm}  
	+ {\bf S}_{\rm cm}, \nonumber \\ 
{\bf K} &=& {1\over 2}\{ {\bf R}_{\rm cm}, H \} - { {{\bf S}_{\rm cm}
       \times {\bf P}_{\rm cm} }\over {H + \M(\k_1,\k_2,\ldots)  }}, 
\label{Def:Kindf}
\end{eqnarray}
with the constraints,
\begin{eqnarray}
& & \sum_a m_a {\bf r}_a = 0, \nonumber \\ 
& & \sum_a {\bf k}_a = 0, \nonumber \\ 
& & {\bf S}_{\rm cm} - \left(
 \sum_a {\bf r_a}\times {\bf k}_a + {\bf s}'_a \right) = 0. 
\end{eqnarray}
In \Eq{Def:Kindf}, the quantity $\M=\M(\k_1,\k_2,\ldots)$ is referred to as
the {\em free} invariant mass in the Schr\"{o}dinger picture.   
The internal momenta ${\bf k}_a$ are related to the
individual particle momenta ${\bf p}_a$ via a free Lorentz transformation
to the CM frame,  
\begin{eqnarray}
{\bf k}_a &=& \Lambda({\bf k}_a \leftarrow {\bf p}_a){\bf p}_a 
	\nonumber \\
&=& 
 {\bf p}_a + { {{\bf p}_a \cdot {\bf P} } 
\over {{\cal M}({\cal M} + H)} }{\bf P} - { {\E(m_a,\p_a)}\over {\cal
  M}} {\bf P} \label{mom}
\end{eqnarray} 
and the CM frame spins ${\bf s}'_a$ are related to the individual spins
${\s}_a$ via a Wigner rotation corresponding to the product of Lorentz
boosts $R=\Lambda(0\leftarrow {\bf p}_a)\Lambda({\bf p}_a \leftarrow 
	{\bf k}_a)\Lambda({\bf k}_a \leftarrow 0)$, leading to
\begin{equation}
{\bf s}'_a =  D^{(s)}(R) {\bf s}_a D^{(s)}(R)^* . \label{spin}
\end{equation}

Interactions are incorporated into the Poincare generators by the
addition of a term in the free mass operator, 
\beq
{\M} \rightarrow \M_I({\bf r}_a,{\k}_a,{\bf s'}_a) = {\M} + V .
\label{Def:Hamdef}
\eeq
Thus, transforming the free Hamiltonian $H$ into the interacting
Hamiltonian $H_I$, 
\bea
H &\to& H_I = H +W, \nonumber \\
W &=& \sqrt{ {\M}_I^2 + {\P}^2} - \sqrt{{\M}^2 + {\P}^2 } .
\eea
This replacement preserves the canonical commutation relations,  
provided $V=V({\bf r}_a,{\bf k}_a,{\bf s'}_a)$ is a function of internal
coordinates only and is invariant under rotations 
$[V,{\bf J}]=[V,{\bf S}_{\rm cm}]=0$. 
For example, consider the case for which the elementary interaction is a
Yukawa-type three-meson vertex.
The matrix elements of the three-meson interaction vertex would be given by 
$\bra{a}V\ket{bc}$ and would only depend on the internal variables
associated with the CM frame where $\p_a=\p_b+\p_c=0$. 
Of course, these internal variables can be expressed in terms of the
individual particle momenta in another frame by using the boost relations
analogous to \Eqs{mom} and \rf{spin}.
 
Within this framework, the Lorentz covariance of observables may be
demonstrated from the following considerations.  
Construct an invariant ${\cal T}$-matrix which satisfies a {\em
Lorentz-invariant} Lippmann-Schwinger equation (LSE),  
\begin{equation}
 {\cal T} = {\cal V} + {\cal V} \, {\cal G} \, {\cal T}, \label{ls0}
\end{equation}
with an invariant interaction ${\cal V}$, 
\bea
 {\cal V} &=& H^2_I - H^2  \nonumber \\
	&=& ( {\bf P}^2 + {\cal M}_I^2 ) 
	  - (  {\bf P}^2 + {\cal M}^2 ) \nonumber \\ 
  &=& W^2 + HW + WH  \nonumber \\
  &=& V^2 + \M V + V \M ,
\eea
and an invariant propagator ${\cal G}$ for scattering energy $E$, given by 
\begin{eqnarray}
{\cal G} &=& ( E^2 - H^2 + i\epsilon)^{-1}  \nonumber \\ 
&=& ( E^2 - {\bf P}^2 - {\cal M}^2 + i\epsilon)^{-1}. 
\label{InvPropagator}
\end{eqnarray}  
Since ${\cal V}$ is independent of the CM momentum $\P$ and the scattering
energy $E$, one may rewrite the scattering energy $E= \sqrt{s + {\P}^2}$,
in terms of a new variable $s$, referred to as the invariant
mass squared.   
From \Eq{InvPropagator}, one observes that the propagator ${\cal G}(s)$ is
a function of the invariant mass squared only, and one concludes that the
LSE \rf{ls0} depends only on the invariant mass
squared $s$.   It follows that the resulting ${\cal T}$-matrix, 
${\cal T}(\sqrt{s})$ depends only on the invariant mass squared $s$. 

It is possible to relate the {\em on-shell} matrix elements of this
invariant ${\cal T}$-matrix to the {\em on-shell} matrix elements of a
$T$-matrix that is the solution of a non-invariant LSE with the
interaction potential $W$, 
\begin{equation}
T(E,\P) = W + W \; G(E) \; T(E,\P),
\end{equation}
where $G(E)=(E-H+i\epsilon)^{-1}$.  The relation between the on-shell
matrix elements is given by 
\begin{equation}
{{\cal T}}(\sqrt{s}) = 2\sqrt{s + {\bf P}^2}\;  T(E,{\bf P}),
\end{equation}
which can be demonstrated term by term by expanding the 
{\em on-shell} matrix elements of  ${\cal T}$ in powers of the potential
$V$,  
\begin{eqnarray}
{\cal T} &=&  {\cal V} + {\cal V}{\cal G}(E){\cal V}  + O({V}^3) 
	\nonumber \\
	&=& W^2 + 2E W \nonumber \\
	&&+ W(E+H) {1\over {E^2\-H^2 \+ i\epsilon} } (E+H) W 
	 + O(V^3)
	\nonumber \\
&=&  2E W  +  2E W {1\over {E \- H \+ i\epsilon} } W + O(V^3) 
	\nonumber \\
  &=& 2E \, T(E,\P) .  \nonumber 
\end{eqnarray}
In this article, calculations are carried out in the CM frame for which
$E=\sqrt{s}$, then the interaction potential $W={\cal M}_I-{\cal M}=V$,
and the relevant LSE is 
\begin{eqnarray}
T(E) &=& V \, + \, V \, G(E) \, T(E), 
	\label{LSE} 
\end{eqnarray}
where $T(E)=T(E,\P=0)$.  In the CM frame, one finds
\begin{eqnarray}
{\cal T}(\sqrt{s}) &=& 2\sqrt{s} \, T(\sqrt{s}) .  \label{TTrel}
\end{eqnarray}
Thus, the on-shell matrix elements of the solution $T(E=\sqrt{s})$ of the
non-invariant LSE in \Eq{LSE} are related by \Eq{TTrel} to the on-shell
matrix elements of the solution ${\cal T}(\sqrt{s})$ of the invariant LSE
of \Eq{ls0}. 
It follows that observables calculated from \Eq{LSE} are equivalent to
those calculated from a Lorentz-invariant theory.

%---------------------------------------------------------------------
\subsection{Coupled Lippmann-Schwinger Equations}
\label{Sec:LSE}
In the above framework, the particle dynamics are given in the
center-of-momentum (CM) frame where ${\bf P}=0$ by the invariant mass
operator, 
\beq
\M_{I} = \M + V \label{M=M0+V}
.
\eeq
The quantity $\M$, introduced in \Eq{Def:Kindf}, is the {\em free}
invariant mass in the Schr\"{o}dinger picture and $V$ is the elementary
hadron interaction potential.   

The probability amplitude for observing an $N$-body state $\ket{\beta\Q}$
with total momentum $\Q$, given an initial $N$-body state $\ket{\alpha\P}$
with total momentum $\P$, is given by the $S$-matrix element
$\bra{\beta\Q}S(E)\ket{\alpha\P}$.    
The $T$-matrix $T(E,\P)$ is defined by the Lippmann-Schwinger equation
(LSE) of \Eq{LSE} and determines the on-shell $S$-matrix
elements,  
\bea
\lefteqn{
\bra{\beta\Q}S(E)\ket{\alpha\P}  = \bracket{\beta\Q}{\alpha\P}
} \nonumber \\ & &- 2\pi i \; \delta(\E(\M_{\beta},\Q) 
 - \E(\M_{\alpha},\P)) \bra{\beta\Q} T(E,\P) \ket{\alpha\P}  
\label{Def:TMatrix},
\eea
where $\E(\M_{\alpha},\P)=\sqrt{\M_{\alpha}^2 + \P^2}$.

The potential $V$ and the $T$-matrix describe all interactions 
between the various channels, including channels with differing numbers of
particles.  In general, they do not conserve particle number.  
Therefore, the LSE of \Eq{LSE} represents an countably infinite system of 
coupled-channel equations which couple states of different numbers of 
particles.

This infinite system of coupled equations may be simplified by
truncating the Hilbert space to include only a finite number of states
that are expected to contribute substantially to a given reaction. 
For the purposes of this study, the Hilbert space is restricted to contain 
a finite number of one-, two-, and three-particle states. 
Furthermore, the particles are assumed to be of finite spatial extension,
thereby providing an ultraviolet regularization to the theory.
With these restrictions, the LSE of \Eq{LSE} reduces to a closed system of
integral equations which may be solved exactly.   
Of course, one drawback of such a truncation is that some symmetries, such
as crossing symmetry, which require the inclusion of many-particle states
may be lost.  
The addition of states with a higher number of particles, such as
four-particle states, can in principle be included in a
straightforward manner but the resulting set of equations would be 
far more complicated than that studied here. 

The main objective of this work is to develop a framework for handling
up to three-body channels in a fully unitary fashion, by including
effects beyond their contribution to the real part of the effective,
two-body potentials.  
The intended application is the description of soft final-state interactions
in hadron production processes.   
Such processes are distinguished by their strong couplings and 
low momentum transfers.  For this reason, composite hadrons (mesons and/or
baryons) are chosen as the fundamental degrees of freedom rather than
quarks and gluons.

The truncation of the Hilbert space to contain only one-, two- and
three-hadron states may be sufficient since, in many applications, states 
with higher numbers of hadrons contribute very little to two-hadron elastic
scattering amplitudes.
This suppression arises because many-hadron intermediate states typically
have a large invariant mass, which appears in the denominator of the Green
function $G$, tending to weaken its contribution. 
Interestingly, this suppression of higher-order Hilbert space states is 
also observed in some quantum field theoretical frameworks.
In a study of the pion-loop contribtion to the $\rho$-meson self-energy
and charge radius, based on a phenomenological application of the
Dyson-Schwinger equations of QCD~\cite{Pichowsky99},  
the covariant, quantum field theoretic expression for the $\rho$-meson
self-energy was separated into the various time orderings and their
relative importance calculated.
The time orderings include contributions arising from $\pi\pi$ and
$\rho\rho\pi\pi$ intermediate states, as well as others.
In this calculation, it was shown that terms associated with the
two-pion intermediate state contributed more than 95\% of the total, 
while the four-hadron states contributed less than 5\%.   
Thus, one expects that a truncation scheme which neglects states with
four or more hadrons should provide a reasonable description of 
the residual strong interactions between mesons and baryons.

The matrix elements of the potential $V$ describe the couplings between
hadrons that arise from the underlying QCD dynamics of quarks and gluons.  
Color confinement requires that all physical particle thresholds are 
associated with the colorless hadron states.  It follows that the matrix
elements of the potential $V$ are real.  
In this framework, all of the analytic structure of the $T$-matrix
necessarily arises from the color-singlet hadron poles and branch cuts
which result from the LSE of \Eq{LSE}. 

Once the Hilbert space has been truncated to include only one-, two- and
three-particle states, \Eq{LSE} is expanded and rewritten in a
simpler form by labeling each of the Hilbert space operators with
subscripts indicating the numbers of particles they act on.
The potential $V$ is of the form
\begin{eqnarray}
V = \left( \begin{array}{ccc}
        V_{11}  & V_{12} & V_{13} \\
        V_{21}  & V_{22} & V_{23} \\
        V_{31}  & V_{32} & V_{33}  \\
        \end{array}
        \right).        \label{VMatrix1}
\end{eqnarray}  
The part of the potential associated with the coupling of a
one-particle state to a two-particle final state is denoted $V_{21}$.  
The resulting system of integral equations can be solved formally 
in a straightforward manner. 

It is important to note that each matrix element of the potential 
$V$ in \Eq{VMatrix1} is itself a matrix, since it 
may contain interactions between any number of different particle 
channels.  
That is, matrix elements of the form $V_{21}$ describe the couplings of
any one-particle state with any two-particle state.   The number of one-,
two-, and three-particle states one wishes to include depends on the
specific application.
%....................................................................
\begin{figure}
\epsfig{figure=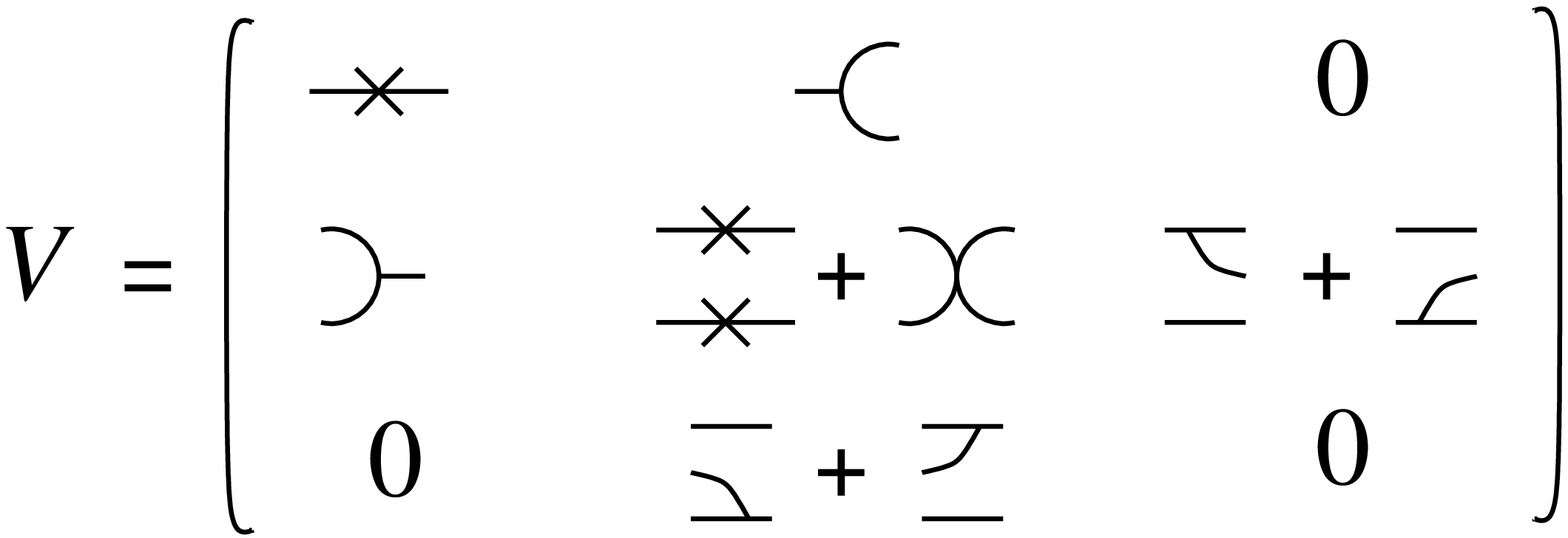,width=8.5cm}
\vspace{0.3truecm}
\caption{Schematic diagram of the interaction potential matrix $V$ of 
\protect\Eq{VMatrix}. }
\label{figone}
\end{figure}
%....................................................................
For the application to $\pi\pi$ scattering considered in \Sec{Sec:PiPi}, a
further simplification is made by assuming an absence of fundamental
interactions in $V$ connecting one-particle states to three-particle
states, and three-particle states to three-particle states.   
Then, the potential $V$ takes a simpler form,
\begin{eqnarray}
V = \left( \begin{array}{ccc}
        V_{11}  & V_{12} & 0 \\
        V_{21}  & V_{22} & V_{23} \\
         0      & V_{32} & 0  \\
        \end{array}
        \right)
, \label{VMatrix}
\end{eqnarray}
and is shown schematically in \Fig{figone}. 

The neglected terms $V_{13} = V_{31}^{\dagger}$ are associated with
energy-independent transitions between one-particle and three-particle
states. When such terms are neglected the only way in which a one-body
state can decay into a  three-body state is through a multiple-step
process involving a two-body intermediate state.  

In setting the term $V_{33}=0$, several possible elementary interactions
have been neglected.
First, $V_{33}$ describes direct energy-independent couplings between two
three-body states, as well as interactions in which two of the particles
interact while the third particle is a spectator. 
Such terms may be important. 
One might argue that it is inconsistent to include direct two-body
interactions in $V_{22}$, but neglect the analogous two-body (plus
spectator) interactions in $V_{33}$. Nonetheless, in this work such 
terms are ignored.  The significance and role of these interactions 
will be addressed in future studies. 

In the truncated Hilbert space, the free Green function is a diagonal matrix
\beq
G = \left(
\begin{array}{ccc}
        G_{1} & 0 & 0 \\
        0   & G_{2} & 0 \\
        0   & 0 & G_{3}  \\
        \end{array}
\right),   \label{GMatrix}
\eeq
and the $T$-matrix is 
\begin{eqnarray}
T = \left( \begin{array}{ccc}
        T_{11}  & T_{12} & T_{13} \\
        T_{21}  & T_{22} & T_{23} \\
        T_{31}  & T_{32} & T_{33}  \\
        \end{array}
        \right).\label{TMatrix}
\end{eqnarray}
In the CM frame, each submatrix $G_{1}$, $G_{2}$ or $G_{3}$ in
\Eq{GMatrix} is itself diagonal since our hadron states form a complete,
orthogonal set of eigenstates of the free invariant mass operator $\M$.

%.................................................................
\begin{figure}
\hspace*{+0.10cm}\epsfig{figure=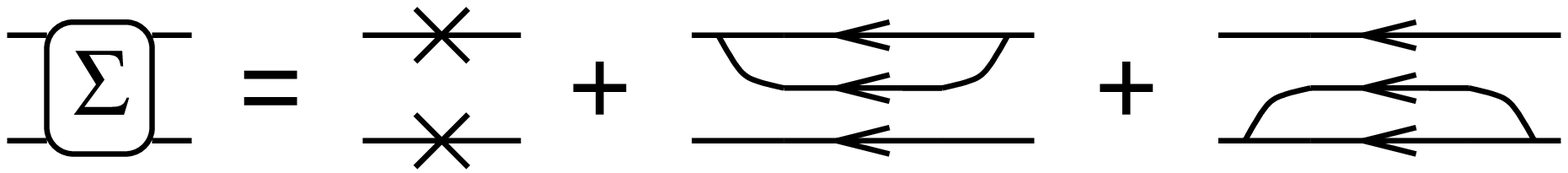,width=8.5cm}\\
\vspace{0.25truecm}
\epsfig{figure=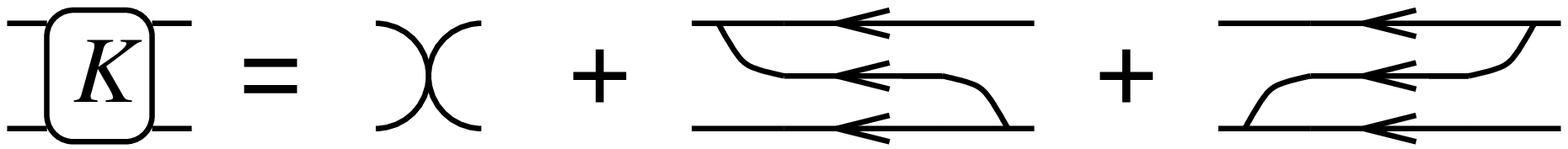,width=8.5cm}
\caption{Schematic diagram of the two-particle self-energy $\Sigma$, 
and the two-particle kernel $K$, as defined in \protect\Eq{Def:SigmaK}. }
\label{figthr}
\end{figure}
%.................................................................
Upon insertion of these forms for $V$, $G$ and $T$ from \Eqs{VMatrix},
\rf{GMatrix} and \rf{TMatrix} into the LSE  \rf{LSE}, one may formally 
solve this system of integral equations.
It is convenient to consider the combination of terms $V_{22} + V_{23}
G_{3} V_{32}$, which appears frequently in our formalism.
These terms play an important role and so are collected and rewritten as
the sum of $\Sigma$ and $K$, 
\begin{eqnarray}
\Sigma + K \equiv V_{22} +  V_{23} G_{3} V_{32} 
        \label{Def:SigmaK}.
\end{eqnarray}
These are referred to as the {\em two-particle self-energy} $\Sigma$, 
and the {\em two-particle kernel} $K$. 
These terms are defined such that the matrix elements of the
two-particle self-energy $\Sigma$ contain only terms proportional to
a $\delta$-function in the {\em relative} momentum of the two-particle
state.   Consequently, matrix elements of the two-particle kernel $K$
contain all contributions that are {\em not} proportional to a
$\delta$-function in the relative momentum.      

The two-particle self-energy and kernel are depicted schematically 
in \Fig{figthr}. 
In the following, it will become apparent that $\Sigma$ and $K$ are the
central elements of the framework, from which all other quantities are
obtained.
In fact, {\em all} effects due to three-particle
intermediate states can be traced back to these two amplitudes.

\begin{figure}[t]
\epsfig{figure=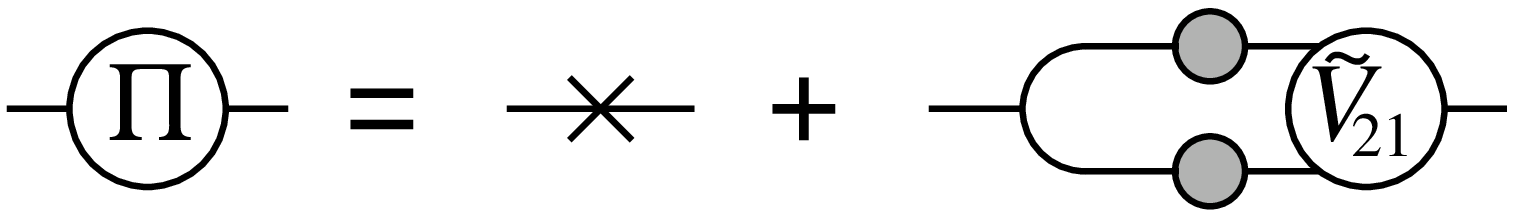,width=8.22cm}\\
\vspace{0.2truecm}
\epsfig{figure=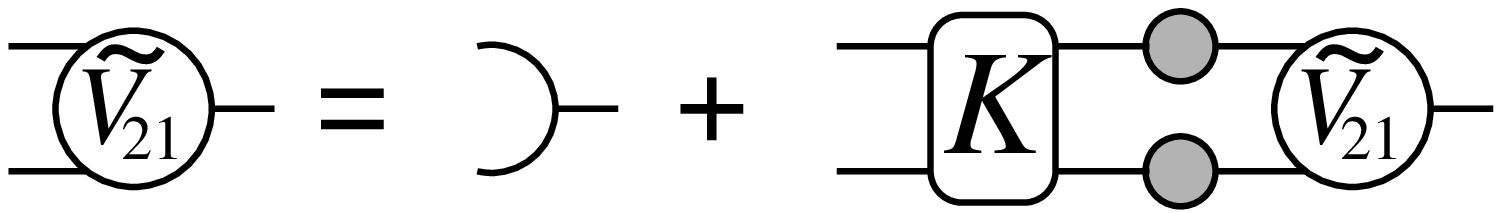,width=8.35cm}
\caption{Schematic diagram of the integral equations for 
the one-body self-energy $\Pi$ and the dressed vertex $\widetilde{V}_{21}$. 
These diagrams depict the expressions in \protect\rf{Def:Pi} and 
\protect\rf{Def:V21}. }
\label{figfiv}
\end{figure}

One defines the {\em dressed} one- and two-particle Green functions in the
usual manner as
\begin{eqnarray}
   \widetilde{G}_{1} &=& ( G_{1}^{-1} - \Pi )^{-1} \label{Def:Gone} ,  \\
   \widetilde{G}_{2} &=& ( G_{2}^{-1} - \Sigma )^{-1} . \label{Def:G2tilde}
\end{eqnarray}
They are defined in terms of the two-body self-energy $\Sigma$ and the
one-body self-energy $\Pi$, where
\begin{eqnarray}
\Pi &=& V_{11} + V_{12} \widetilde{G}_{2} \widetilde{V}_{21} \label{Def:Pi}, \\
\widetilde{V}_{21}&=& V_{21} + K \widetilde{G}_{2} \widetilde{V}_{21},
	\nonumber \\
	&=& ( 1 - K \widetilde{G}_{2} )^{-1} V_{21}. \label{Def:V21}
\end{eqnarray}
These quantities are shown schematically in \Fig{figfiv}.

%...............................................................
%...............................................................

\begin{figure}
\epsfig{figure=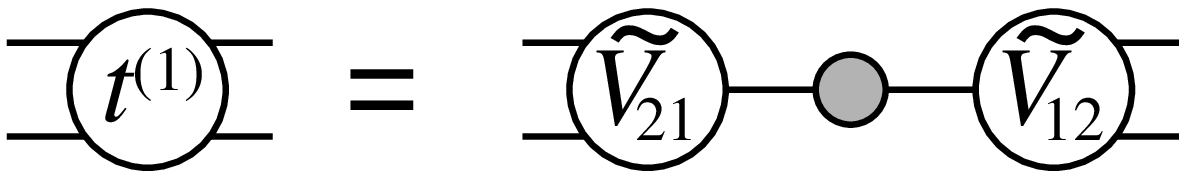,width=6.35cm}\\
\vspace{0.2truecm}
\epsfig{figure=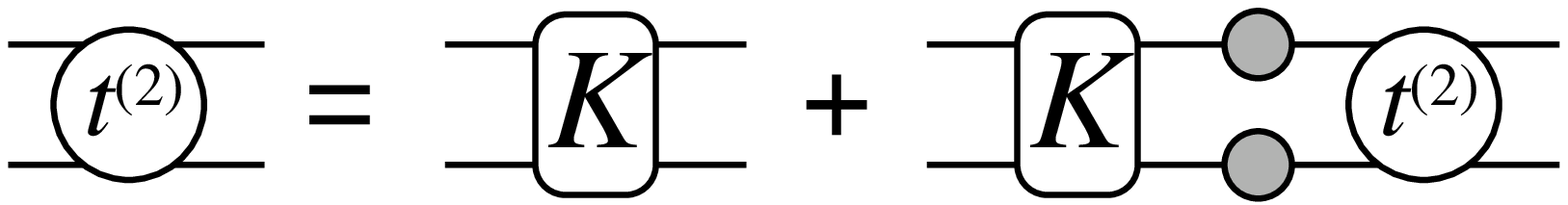,width=8.5cm}
\caption{Schematic diagram of the scattering amplitudes $t^{(1)}$ 
and $t^{(2)}$ which enter into the two-particle scattering matrix
elements. 
These diagrams depict the expressions in 
\protect\Eqs{Def:t1} and \protect\rf{Def:t2}.}
\label{figsvn}
\end{figure}

The solution for the two-particle scattering $T$-matrix 
can be written as 
\begin{eqnarray}
T_{22} &=& G^{-1}_{2} \widetilde{G}_{2} (t^{(1)} + t^{(2)} ) 
  \widetilde{G}_{2} G^{-1}_{2} + G^{-1}_{2} \widetilde{G}_{2} \Sigma ~, 
\label{LSE22}
\end{eqnarray}
where
\begin{eqnarray}
t^{(1)} &=& \widetilde{V}_{21} \widetilde{G}_{1} \widetilde{V}_{12} ,
        \label{Def:t1}\\
t^{(2)} &=& K + K \, \widetilde{G}_{2} \, t^{(2)}, \nonumber \\
	&=& (1 - K \widetilde{G}_{2})^{-1} K.
	\label{Def:t2}
\end{eqnarray}
These two scattering amplitudes are shown schematically in Fig.\ 
\ref{figsvn}.  Briefly, the contributions to the two-body scattering
amplitude $T_{22}$ that proceed through one-body channels are denoted
$t^{(1)}$, while contributions that don't proceed through one-body
channels are denoted $t^{(2)}$;  
both $t^{(1)}$ and $t^{(2)}$ contain the effects of the two- and
three-body singularities, but only $t^{(1)}$ contains one-body
singularities.  
%...........................................
%............................................

%------------------------------------------------------------------------
The matrix elements for the {\em dressed} Green function
$\widetilde{G}_{2}$ defined by \Eq{Def:G2tilde} are given by
\beq
\widetilde{G}_{\beta\alpha}(p,E)  = 
\bigg( \delta_{\alpha\beta}(E \- \M_{\alpha_{12}}(p) \+ i \epsilon)
 \- \Sigma_{\alpha\beta}(p,E) \bigg)^{-1}.
\eeq
One can collect the terms from \Eq{Def:SigmaK} contributing to the two-body
self-energy $\Sigma_{\beta\alpha}$, and organize them into the following
sum,
\beq
\Sigma = \delta\Sigma
+ V^{(1)}_{23}G_{3}V^{(1)}_{32} + V^{(2)}_{23}G_{3}V^{(2)}_{32}.
\label{Def:Sigma}
\eeq
Here, the superscript $(i)$ refers to the diagram in which the 
$i^{\rm th}$ particle in the two-body state emits and subsequently 
re-absorbs the particle $\gamma_3$.
The term $\delta\Sigma$ is identified with the part of the potential 
$V_{22}$ that is proportional to a Dirac $\delta$-function in the relative
two-body momentum.
(All other terms that appear in \Eq{Def:SigmaK} but which do not
appear in $\Sigma$ in \Eq{Def:Sigma}, are part of the two-body kernel
$K$.)   

\begin{figure}
\center{\epsfig{figure=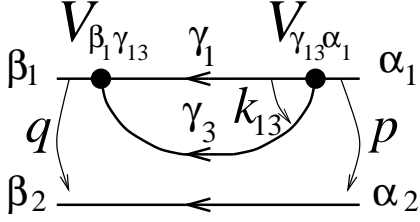,width=5.50cm}}
\vspace*{0.350cm}
\caption{Diagram depicting one of the contributions to the two-body
self-energy $\Sigma_{\beta\alpha}^{(1)}(p,E)$ given in \Eq{Sigma1Int}. 
Shown here, particle $\alpha_1$ decays into particles $\gamma_{1}$ and
$\gamma_{3}$, and these subsequently recombine to form particle
$\beta_{1}$.
The relative momentum between the intermediate particles
$\gamma_{1}$ and $\gamma_{3}$ is $k_{13}$, 
the relative momentum of the incoming state $\ket{\alpha\p_1\p_2}$ is $p$,
and relative momentum of the outgoing state $\ket{\beta\q_1\q_2}$ is $q$.
Solid circles denote matrix elements of the potentials from 
\Eq{VMatrix}, $V_{32}$ and $V_{23}$, evaluated between two- and
three-body states.
}\label{Fig:Sigma1}
\end{figure}

Upon inserting a complete set of three-body states into \Eq{Def:Sigma}
and evaluating the resulting expressions in the overall CM frame with
$\P=0$, one obtains
\bea
\lefteqn{
        \Sigma_{\beta\alpha}^{(1)}(p,E)=  
        \sum_{\gamma_{13}}\int_{0}^{\infty}\!\!\!\!dk_{13} \; 
        \frac{a_{\gamma_{13}}(k_{13},p)}
        {2 \sqrt{ \E(m_{\beta_1},p) \E(m_{\alpha_1},p) } } \;
} \nonumber \\ && \times
V_{\beta_{1}\gamma_{13}}(k_{13}) \;\;
G_{\gamma}(p,k_{13},E) \; \;
V_{\gamma_{13}\alpha_{1}}(k_{13}) 
, \label{Sigma1Int}
\eea
where $\Sigma^{(i)}=V^{(i)}_{23}G_{3}V^{(i)}_{32}$. 
A similar expression is obtained for $\Sigma^{(2)}_{\beta\alpha}(p,E)$.
The momentum integration is over the relative momentum $k_{13}$ between
the first and third particles of the three-body intermediate state, $J$ is
the total angular momentum of the system,  the sum is over all three-body
states $\gamma$, and the three-body Green function is  
\beq
G_{\gamma}(p,k_{13},E) = 
\frac{1}{E - \M_{\gamma_{123}}(p,k_{13}) + i\epsilon}.
\label{Def:GF3body}
\eeq
For brevity the ubiquitous two-body phase space factor, 
\beq
a_{\gamma_{13}}(k_{13},p) = \frac{k_{13}^2}{(2\pi)^3} 
        \frac{\rho_{\gamma_{13}}(k_{13})}
        {2 \E(\M_{\gamma_{13}}(k_{13}),p)}
,
\label{eq:twophase}
\eeq
and two-body Jacobian 
\beq
\rho_{\gamma_{13}}(k_{13})= \frac{\M_{\gamma_{13}}(k_{13})}
	{2 \E(m_{\gamma_1},k_{13}) \E(m_{\gamma_3},k_{13}) },
\eeq
are introduced.   
The expression in \Eq{Sigma1Int} for the two-body self-energy
$\Sigma^{(1)}_{\beta\alpha}(p,E)$ is depicted in \Fig{Fig:Sigma1}.
The two-body self-energy $\Sigma_{\beta\alpha}(p,E)$ is
then the sum, 
\beq
\Sigma_{\beta\alpha}(p,E) = \delta\Sigma_{\beta\alpha}(p) + 
\Sigma^{(1)}_{\beta\alpha}(p,E) + \Sigma^{(2)}_{\beta\alpha}(p,E) ,
        \label{SigmaSum}
\eeq
where the counter term is chosen to be
\beq
\delta\Sigma_{\alpha}(p) 
= - \bigg( \Sigma_{\alpha}^{(1)}(p,E)
+ \Sigma_{\alpha}^{(2)}(p,E) \bigg)_{E = \M_{\alpha_{12}}(p)}
        . \label{deltaSigma}
\eeq
This is necessary and sufficient to ensure unitarity and 
that the stable two-body system $\alpha_{12}$ is observed asymptotically
with the invariant mass $\M_{\alpha_{12}}(p)$.  
Evaluation of the two-body self-energy $\Sigma_{\alpha}(p,E)$ requires
calculating the imaginary part and a principal part of the integral in
\Eq{Sigma1Int}.
For energies $E$ above a three-body threshold, these integrals 
encounter poles in the three-body Green function
$G_{\gamma}(\p,\k_{13},E)$ for values of the relative momentum 
$k_{13}=k_{0\gamma}$, where $k_{0\gamma}$ satisfies the relation 
$\M_{\gamma_{123}}(p,k_{0\gamma})=E.$ 

%.............................................
%.............................................

{From} \Eq{Def:Pi}, one obtains an expression for the one-body
self-energy in the CM frame,
\bea
   \Pi_{\alpha}(E) &=& \delta\Pi_{\alpha}
   + \frac{1}{2 m_{\alpha_1} } \sum_{\gamma_{12}}
   \int_{0}^{\infty}\!\!\!\! dk  \;\; a_{\gamma_{12}}(k,0) \; 
  \nonumber \\
& &\times 
        V_{\alpha_1 \gamma_{12}}(k) \; 
        \widetilde{G}_{\gamma}(k,E) \;
        \widetilde{V}_{\gamma_{12}\alpha_{1}}(k,E) \; . 
\label{PiIntegral}
\eea
In \Eq{PiIntegral}, $m_{\alpha_1}$ is the mass of the one-body state
$\ket{\alpha\P}$,  
$V_{\alpha_{1}\gamma_{12}}(k)$ is the vertex function of the potential
$V_{12}$, 
$\widetilde{V}_{\gamma_{12}\alpha_{1}}(k,E)$ is the vertex function for 
the dressed vertex $\widetilde{V}_{21}$, 
$\widetilde{G}_{\gamma}(k,E)$ is the dressed two-body Green function,  
and $a_{\gamma_{12}}(k,0)$ is a factor from \Eq{eq:twophase} 
associated with the phase space of the two-body system $\gamma_{12}$.  
The one-body mass counter term
$\delta\Pi_{\alpha}$ is fixed by demanding that the elements of the
one-body self-energies be identically zero when the driving energy 
$E = \E(m_{\alpha_1},\P)$.  In the CM frame, the mass 
renormalization condition is 
\beq
{\rm Re}
\big( \Pi_{\alpha}(E) \big)_{E\rightarrow m_{\alpha_1}} = 0 ~.
        \label{deltaPiConstraint}
\eeq
In this framework, the finite size of the hadrons involved results in
vertex form factors, such as ${V}_{\beta_{1}\gamma_{12}}(k)$, which fall
off sufficiently rapidly with $k$ to ensure the convergence of all
integrals. 
Therefore, the counter terms $\delta\Pi$ and $\delta\Sigma$ are both
finite.

Having obtained expressions for the dressed one- and two-boby Green
functions, one next considers the two-body scattering amplitudes $t^{(1)}$
and $t^{(2)}$.  These scattering amplitudes depend on the one- and
two-body Green functions, as well as the two-body kernel $K$.
The two-body kernel is comprised of three contributions
\bea
K &=& K^{(1)} + K^{(2)} +  K^{({\rm 4pt})},
\label{K=K1+K2+K3}
\eea
where 
\bea
K^{(1)} &=& V_{23}^{(2)} G_{3} V_{32}^{(1)}, \label{K1=VGV}\\
K^{(2)} &=& V_{23}^{(1)} G_{3} V_{32}^{(2)}, \label{K2=VGV}
\eea
and $K^{({\rm 4pt})}$ is the part of the potential $V_{22}$ that is 
{\em not} proportional to a Dirac $\delta$-function in the relative
momentum.  This latter term is the direct four-point coupling of four
mesons, and it is depicted as four meson lines converging on a single
point in \Fig{figone}. 
In \Eq{K=K1+K2+K3}, the parenthetical superscripts on the first two
quantities refer to which of the two particles in the incoming state emits
the exchanged particle, the superscript on the third quantity refers to
the direct four-meson interaction.   
In the first two terms, the exchanged particle is subsequently absorbed by
the other particle in the outgoing state. 
%.............................................
\begin{figure}
\center{\epsfig{figure=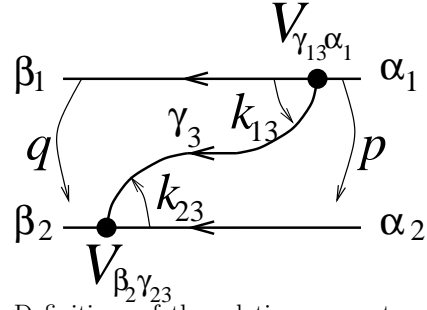,width=5.50cm}}
\caption{Definitions of the relative momenta and particle labels for the
kernel $K_{\beta\alpha}^{(1)}(q,p)$ given in \protect\Eq{ExplicitK1J}.}
\label{Fig:K1}
\end{figure}
%.............................................

Explicit expressions for the most general two-body kernels of \Eqs{K1=VGV}
and \rf{K2=VGV} in the spherical wave basis  are complicated and not
particularly enlightening.  
However, in the model application to $\pi\pi$ scattering considered in
\Sec{Sec:PiPi}, the resulting kernel is relatively simple.  The only
matrix elements of $V_{23}$ that are of interest in this application are
those associated with the transitions of the forms 
$\pi\pi\rightarrow\pi\pi\rho$, $\pi\pi\rightarrow\pi\pi f_0$, 
$\pi\pi\rightarrow K \bar{K} \rho$ and $\pi\pi\rightarrow K \bar{K}f_0$, 
$\pi\pi\rightarrow \pi \bar{K} K^*$.
In each of these hadron states, at least two of the three particles are
spin-0 mesons.  
For these interactions, the plane-wave matrix elements of the potentials
$V^{(1)}_{23}$ are of the form 
\bea
\bra{\gamma_{123} \K \k \k_{13} } V^{(1)}_{32} \ket{\alpha_{12} \P \p}&=&
	(2\pi)^3\delta(\K-\P) \delta_{\gamma_{2},\alpha_{2}} 
	\nonumber \\
 && \times 
 	2 \E(m_{\alpha_{2}},-\p)
 	\nonumber \\
 && \times
	\bra{\gamma_{13} \k \k_{13} } V_{21} \ket{\alpha_{1} \p},
\eea
where the matrix elements of $V_{12}$ are of the form
\bea
\lefteqn{
\bra{\gamma_{13} \k \k_{13} } V_{21} \ket{\alpha_{1} \p}
= (2\pi)^{3}\delta(\k-\p)
	\frac{(-1)^{s_{\gamma_3}-\lambda}}{\sqrt{2 s_{\gamma_3} + 1}} 
} \nonumber \\
&&      \sum_{\lambda}
        D^{s_{\gamma_3}}_{\lambda_{\gamma_3},\lambda}(-\k_{13},\p) 
        \; k_{13}^{s_{\gamma_3}}
        \; Y^*_{s_{\gamma_3},-\lambda}(\hat{\k}_{13}) 
        \; V_{\gamma_{13}\alpha_{1}}(k_{13}).
        \label{V21:GenSpin}
\eea
The vertex for $V_{32}^{(1)}$ appears in Figs.~\ref{Fig:Sigma1} and
\ref{Fig:K1} as the right-most interaction vetrtex.
In the partial-wave basis, 
the kernel $K^{(1)}$ is shown in \Fig{Fig:K1} and is given by 
\bea
K^{(1)J}_{\beta\alpha}(q,p)&=&
\! 2\pi \!
\sum_{\gamma_{123}} 
\int_{-1}^{+1} \!\!\! dx \frac{P_{J}(x)}{4\pi} 
        \frac{1}{2\E(m_{\gamma_3},\p-\q)}
\nonumber \\
&&\times 
        \frac{  V_{\beta_{2}\gamma_{23}}(k_{23})
                \; {\cal S}^{s_{\gamma_3}}(\q,\p) \;
                V_{\gamma_{13}\alpha_{1}}(k_{13}) 
        }{E - \M_{\gamma_{123}}(\q,-\p,\p-\q) + i\epsilon}
        ,       \label{ExplicitK1J}
\eea
where 
\bea
S^{s_{\gamma_{3}} = 0 }(\q,\p)&=& 1, \label{S:Scalar}  \nonumber \\
S^{s_{\gamma_{3}} = 1 }(\q,\p)&=&  q_{\beta_1}^{\mu} \; 
\left(
\frac{-m_{\gamma_{3}}^2g_{\mu\nu}+k_{\gamma_3\mu}k_{\gamma_3\nu}}
{\M_{\gamma_{23}}(k_{23})\M_{\gamma_{13}}(k_{13})} \right) 
         p_{\alpha_2}^{\nu} \label{S:Vector}, \!\!
\eea
for scalar exchanges ($s_{\gamma_3}=0$) and vector exchanges
($s_{\gamma_3}=1$), respectively, and $P_{J}(x)$ are the usual Legendre
polynomials in $x=\p\cdot\q/pq$.  
The three four-momenta appearing in \Eq{ExplicitK1J} are
\bea
q_{\beta_1}^{\mu}&=& \left(\E(m_{\beta_{1}},\q),\q \right), \nonumber \\
p_{\alpha_2}^{\mu}&=& \left(\E(m_{\alpha_{2}},-\p),-\p \right),\nonumber \\
k_{\gamma_3}^{\mu}&=& \left(\E(m_{\gamma_{3}},\p-\q) , \p-\q \right).
\eea
Expressions corresponding to the matrix elements of
$K^{(2)}_{\beta\alpha}(q,p)$ can be obtained in a similar manner.

Once specific forms of the model vertex form factors
$V_{\gamma_{13}\alpha_{1}}(k_{13})$ and 
$V_{\beta_{2}\gamma_{23}}(k_{23})$ are
provided and substituted into \Eq{ExplicitK1J}, the matrix elements of the
kernel $K$ are computed numerically.
One can proceed to solve the integral equation in \Eq{Def:t2} for the
scattering amplitude $t^{(2)}$.   In the CM frame, the integral equation
for the partial-wave scattering amplitude $t^{(2)J}_{\beta\alpha}(q,p)$
has the form 
\bea
  t^{(2)J}_{\beta\alpha}(q,p) &=&
          K^{J}_{\beta\alpha}(q,p) 
          + 
          \sum_{\gamma_{12}} \int_{0}^{\infty} \!\!\! dk \,
        a_{\gamma_{12}}(k,0)
\nonumber \\
& & \times   
        K^{J}_{\beta\gamma}(q,k) \; \widetilde{G}_{\gamma}(k,E) \;
        t^{(2)J}_{\gamma \alpha}(k,p),
\label{eq:t22sol}
\eea
where $a_{\gamma_{12}}(k,0)$ is the usual two-body phase space factor.
Obtaining the solution of this integral equation is complicated by 
the presence of the two-body pole in the two-body Green function
$\widetilde{G}_{\gamma}(k)$, and possibly the appearance of three-body
unitarity cuts in both $\widetilde{G}_{\gamma}(k)$ and the two-body kernel
$K^{J}_{\beta\alpha}(q,p)$.  The method used to solve this integral
equation is adapted from \Ref{Haftel}.  It involves obtaining 
a two-body Moller operator $\Omega^{(2)}$, whose $J^{\rm th}$ partial-wave
matrix element satisfies, 
\beq
 t^{(2)J}_{\beta\alpha}(q,p)= \sum_{\gamma_{12}} 
        \!\! \int_{0}^{\infty}\!\!\!\!\! dk 
        \; a_{\gamma_{12}}(k,0) \; \Omega^{(2)J}_{\beta\gamma}(q,k) \; 
        K_{\gamma\alpha}(k,p).  \label{t=OmegaK}
\eeq
After a solution for $t^{(2)J}_{\beta\alpha}(q,p)$ is obtained, one 
proceeds to obtain an explicit expression for the two-body scattering
amplitude $t^{(1)}$, which is given by \Eq{Def:t1}.  
Since the solution of the intermediate dressed 
one-body Green function $\widetilde{G}_{1}$ is obtained from the one-body
self-energy from \Eq{PiIntegral}, all that remains is to determine the
form of the dressed vertices $\widetilde{V}_{12}$ and
$\widetilde{V}_{21}$.  
The integral equation for the dressed vertex is obtained for 
the transpose of the dressed vertex from \Eq{Def:V21},
\bea
\widetilde{V}_{\beta_{12}\alpha_{1}}(q)& = &
{V}_{\beta_{12}\alpha_{1}}(q) 
+ \sum_{\gamma_{12}} \int_{0}^{\infty}\!\!dk \; a_{\gamma_{12}}(k,0)
\nonumber \\ & & \times 
        \; K^{J=s_{\alpha_1}}_{\beta\gamma}(q,k) 
        \; \widetilde{G}_{\gamma}(k,E) 
        \; \widetilde{V}_{\gamma_{12}\alpha_{1}}(k) 
\label{eq:Vprime}
\eea
The similarity between this integral equation and the integral equation of  
\Eq{eq:t22sol} with $J=s_{\alpha_1}$ which determines $t^{(2)}$ is
clear.  
It follows that the solution to \Eq{eq:Vprime} is just
\beq
\widetilde{V}_{\beta_{12}\alpha_{1}}(q) 
        = \sum_{\gamma_{12}} 
        \!\! \int_{0}^{\infty}\!\!\!\!\! dk 
        \; a_{\gamma_{12}}(k,0) \; 
      \Omega^{(2)J=s_{\alpha_1}}_{\beta\gamma}(q,k) \; 
        V_{\gamma_{12}\alpha_{1}}(k).
\eeq
Finally, the two-body scattering amplitude $t^{(1)}$ is 
\beq
t^{(1)J}_{\beta\alpha}(q,p) = 
\sum_{\gamma_{1}'\gamma_{1}} 
\frac{\widetilde{V}^{J}_{\beta_{12} \gamma_{1}'}(q)
\widetilde{G}_{\gamma_{1}'\gamma_1}(E)
\widetilde{V}^{J}_{\gamma_{1} \alpha_{12}}(p)}
{2\sqrt{m_{\gamma_{1}'} m_{\gamma_{1}}}}
        \label{t1fin}
\eeq 
The complete expression for the two-body scattering amplitude $T_{22}$ is
obtained by adding this expression for $t^{(1)}$ to $t^{(2)}$ according to
\Eq{LSE22}. 

In the previous sections, it was demonstrated that the explicit solution
to the scattering problem involving one-, two-, and three-body states can 
be obtained by performing several integrations and one matrix inversion.
The matrix inversion is necessary to obtain the two-body Moller amplitude
$\Omega^{(2)J}_{\beta\alpha}(q,p)$.

%------------------------------------------------------------------------
\section{Application to $\pi\pi$ scattering}
\label{Sec:PiPi}
In this section, the framework is applied to $\pi\pi$ scattering.
Simple model forms of the elementary vertex form factors
$V_{\beta_{12}\alpha_{1}}(q)$ are introduced, and solutions for the
self-energies 
$\Pi_{\beta\alpha}(E)$ and $\Sigma_{\beta\alpha}(p,E)$ and scattering
amplitudes $t^{(1)J}_{\beta\alpha}(q,p)$, $t^{(2)J}_{\beta\alpha}(q,p)$, 
and $\widetilde{V}_{\beta_{12}\alpha_{1}}(q)$ are obtained numerically.    
Several interesting aspects of the obtained solutions are discussed.  
It should be emphasized that the model introduced in \Sec{Sec:Model}
is preliminary and the manner in which the model parameters are fit to the
data may be overly simplistic, as it focuses on reproducing only a few
observables and therefore does not represent an exhaustive or complete
study of the dynamics of $\pi\pi$ scattering.  
The motivation is to provide a demonstration of the framework and exhibit
the features of the model, and its ability to describe the scattering of 
a system of
strongly-coupled particles with emphasis on the multiparticle channel aspect. 
More complete studies of meson scattering within the present framework will
be the subject of future articles.

In \Sec{Sec:Model}, the dynamical assumptions are discussed along with the
model parameters.   A detailed list of the states included in the
Hilbert space is provided.
The model parameters are determined using a simple method to 
fit experimental data for the $\pi\pi$ isoscalar-scalar phase shift and 
$\rho$-meson decay width, using the $S$-wave phase shifts from 
\Ref{PiData}.   
In \Sec{Sec:PhiS} the resulting phase shifts, inelasticities and cross
sections are provided and compared to the data, and some aspects of the
$f_{0}(980)$ scalar meson are discussed in terms of a $K\bar{K}$ bound
state.

%--------------------------------------------------------------------------
\subsection{Dynamical model for $\pi\pi$ scattering}
\label{Sec:Model}

The model is intended to describe the scattering in a range
of center-of-momentum (CM) energies from threshold 
($E= 2m_{\pi}\approx$ 280~MeV) to about $E=$ 1400~MeV.  
Above 1400~MeV, it is important to include 
in more detail the effects of the three scalar mesons observed in 
this region.  For this preliminary study, however, it is possible to avoid
making strong assumptions concerning these scalar mesons, hence the model
will not be accurate in this energy region.
In the following, only the isoscalar-scalar $I=0, J=0$) and
isovector-vector ($I=1,J=1$) channels are considered. 
The motivation is to explore some of the interesting physical aspects of the
present framework and to estimate the importance of including three-body
states in such a model of hadron scattering.
The assumptions of the dynamical model are summarized below.

{\em Two-body states:}
 For the channels and energies explored herein, it is assumed that 
$\pi\pi$ scattering is primarily determined by the dynamics arising from
the coupling of the $\pi\pi$ and $K\bar{K}$ two-body channels.
Hence, $\ket{\pi\pi}$ and $\ket{K\bar{K}}$ are the only two-body
channels included in the Hilbert space.

{\em One-body states:}
It is assumed that the coupling of the $K\bar{K}$ system is
strong enough to result in the appearance of a narrow resonance in the
scalar-isoscalar channel at $E\approx$ 980~MeV.  
This state is identified with the $J^{PC}=0^{++}$ $f_0(980)$ meson. 
Since this scalar meson is presumed to arise from final-state interactions
as a quasi-bound $K\bar{K}$ state, it is not part of the free Hilbert 
space, and there is no bare mass associated with it. 
Rather, it appears as a pole in the analytically-continued $T$-matrix.
Furthermore, in the limit that the two-body $\pi\pi$ and $K\bar{K}$
channels decouple, this pole moves to the real-energy axis below the
two-kaon threshold;  that is, it becomes a $K\bar{K}$ bound state 
in this limit.  The identification of the $f_0(980)$ meson as 
a $K\bar{K}$ molecule is controversial.  Although it appears as a 
molecular state in this model, the ``true'' nature of the $f_0(980)$ meson 
remains an open question.  

In contrast to the $f_0(980)$ meson, it is assumed that at least one
of the scalar resonances observed in the mass region between 1300 and
1700~MeV will be a QCD bound state; that is, a state which arises as 
a bound state whose constituents are quarks, antiquarks and gluons.   
Such states do not arise from the meson final-state interactions, they are
not bound states of mesons, and hence must be included in the model
as {\em bare} states with bare masses. 

Experiments reveal the presence of several resonances in the
scalar-isoscalar channel between 1300 and 1700~MeV.   A complete study of
the $\pi\pi$ scattering system in this energy range requires the inclusion of
each of these resonances into the model.   
However, to simplify the present study, all of these resonances are
modeled in terms of a single scalar resonance.
The resonance is assumed to have a mass of 1350~MeV, which gives it a mass 
similar to the lightest of the resonances above kaon threshold,
referred to as the $J^{PC}=0^{++}$ $f_0(1370)$ meson.
One ramification of choosing a single scalar resonance to model the effect
of all observed resonances in the 1300--1700~MeV region, is 
in the width of this resonance.  In order to 
fit the model parameters to the $\pi\pi$ phase shifts requires 
a single effective resonance with a very large width.
It is found that the model resonance has a decay width of
805~MeV, which is approximately the {\em sum} of the widths of the three
observed resonances in this region. 

{\em Three-body states:}
Only three-body states are included that can couple to the $\pi\pi$ or
$K\bar{K}$ states through the absorption or emission of the 
isovector $J^{PC}=1^{--}$ $\rho(770)$, isodoublet 
$J^{P}=1^{-}$ $K^*(892)$,  and $J^{PC} = 0^{++}$ $f_{0}$ mesons.
Thus, the three-body states included in this study are
$\ket{\pi\pi\rho}$, $\ket{\pi\pi f_0}$, 
$\ket{K\bar{K}\rho}$, $\ket{K\bar{K} f_0}$, 
$\ket{\pi\bar{K}K^*}$ and $\ket{\pi{K}K^*}$.  

To summarize, the hadronic states included in this model of $\pi\pi$
scattering are 
\bea
&&\ket{f_0}, \; \ket{\rho},  \nonumber \\ 
&&\ket{\pi\pi}, \; \ket{K\bar{K}}, \nonumber \\
&&\ket{\pi\pi \rho }, \, \ket{K\bar{K}\rho}, \, \ket{\pi\bar{K}K^*}, \,
\ket{\pi\pi f_0}, \, \ket{K\bar{K} f_0},
\label{Channels}
\eea
where the $f_0$ meson refers to the $f_0(1350)$ meson.
(The $f_0(980)$ meson is expected to appear in the model as a $K\bar{K}$
resonance.) 
The values of the bare masses of these particles are provided in
Table~\ref{Tab:Particles} and are underlined to indicate that they are
input parameters.
As discussed in \Sec{Sec:Formalism}, one- and two-body counter terms are
included in the elementary interaction potentials $V_{11}$ and $V_{22}$,
respectively, such that the bare masses given in Table~\ref{Tab:Particles}
coincide with the dressed masses of the mesons.  

%*************************************** Particle properties.
\begin{table}
\caption{
Masses and widths of mesons.  Masses that are underlined have been fixed
to reproduce the accepted values.  All other values are obtained from
the model calculation.
In the present study, the width of the $K^*$ meson was not calculated.
All values are given in units of MeV.} 
\label{Tab:Particles}
\begin{tabular}{ccccccc}
 & $\pi$ & $K$ & $\rho$  & $K^*$ & $f_0(1350)$ & $f_0(980)$ \\ \hline
mass & $\underline{140}$  & $\underline{500}$  & $\underline{770}$  
        & $\underline{890}$  & $\underline{1350}$ & 996 \\
width&  0  & 0 & 150 & ---  & 805 &  46
\end{tabular}
\end{table}
%************************************************************
%*********************************************** Isospin table for K
\begin{table}
\caption{
Isospin coupling constants for kernel $K_{\beta\alpha}(q,p)$ for  
isoscalar, $S$-wave ($I=0$, $J^{PC}=0^{++}$) and 
isovector, $P$-wave ($I=1$, $J^{PC}=1^{--}$) scattering.}
\label{Tab:Kernel}
\begin{tabular}{c |ccc}
Channel & Exchange  & $I=0$, $J=0$ & $I=1$, $J=1$ \\ \hline
$\pi\pi \leftrightarrow \pi\pi$& $\pi\pi f_0$ & 1 & 1\\
$K\bar{K}\leftrightarrow K\bar{K}$& $K\bar{K}f_0$ & 1 & 1\\
$\pi\pi \leftrightarrow \pi\pi$& $\pi\pi\rho$ & -1 & -1 \\
$K\bar{K}\leftrightarrow K\bar{K}$& $K\bar{K}\rho$ & -1 & -1 \\
$\pi\pi\leftrightarrow K\bar{K}$& $\pi{K}K^*$ & $-\sqrt{2}$ & -1 \\
\end{tabular}
\end{table}
%*******************************************************************

%\item  
{\em Model vertex form factors:}
The vertices in the model are assumed to be finite-sized and hence
require the appropriate form factors for the relative three-momentum
$\q$.  They are given by the universal form:  
\beq
V_{\beta_{12}\alpha_{1}}(q) = a_{\beta_{1}\beta_{2}\alpha_{1}}
        \sqrt{\frac{16\pi}{2s_{\rm max}+1}} 
        \; e^{-q^2 / \Lambda_{\beta_1\beta_2\alpha_1}^2 }
        ,
        \label{Vertex3}
\eeq
where $s_{\rm max}={\rm max}\{s_{\beta_1},s_{\beta_2},s_{\alpha_1}\}$ is
the largest spin of the particles involved.  
In the present study $s=0$ for vertices involving the $f_0$ meson, and
$s=1$ for vertices involving the $\rho$ or $K^*$ mesons. 
The vertex coupling constants $a_{\beta_1\beta_2\alpha_1}$ and form
factor momentum scales $\Lambda_{\beta_1\beta_2\alpha_1}$ are chosen to
provide a good fit to the data for the isoscalar-scalar $\pi\pi$ phase
shift $\delta_{\pi\pi}$ and the $\rho$-meson decay width
$\Gamma_{\rho\rightarrow\pi\pi}=150$ MeV.
The parameter search was limited in a number of ways.   
First, the various meson-exchange form factor scales
$\Lambda_{\beta_1\beta_2\alpha_1}$ in Table~\ref{Tab:Couplings} were all
constrained to be the same value and less than 1~GeV. 
The direct four-meson couplings in Table~\ref{Tab:Couplings2} were chosen
to be one of two scales, the first was taken to be 125~MeV larger, and the
second to be 125~MeV smaller than the meson-exchange scales in
Table~\ref{Tab:Couplings}. 
In the following, it is shown that the vector-exchange interactions
contribute little to the observables considered.  
To reduce the number of parameters, the strength of the vector-exchange
vertices were taken to be identically equal
$a_{\pi\pi\rho}=a_{K\bar{K}\rho}=a_{\pi K K^*}$. 
The isospin factors that arise in a calculation of the meson-exchange
kernels $K^{J}_{\beta\alpha}(q,p)$, such as in \Eq{ExplicitK1J}, are given
in Table~\ref{Tab:Kernel}. 

%.......................
{\em Direct interactions:} 
In addition to meson exchanges, it is important to include real-valued
potentials that directly couple two pseudoscalar mesons to two
pseudoscalar mesons, as a part of the $K^{({\rm 4 pt})}$ kernel 
in \Eq{K=K1+K2+K3}.  
Such interaction potentials could arise from the direct coupling of four
mesons to a virtual-quark loop.  Here, two direct four-point interactions
are considered.  The first is intended as a way to mimic some of the effects
of dynamical chiral symmetry breaking.  This interaction is taken to be of
the form given by the elementary potential $V_{22}$ and is referred to as
the direct $4\pi$ (or $4K$) interaction.   In a partial-wave basis, the
form of the four-pion interaction is given by 
\beq
K^{{(4 \pi)}J}_{\pi\pi,\pi\pi}(q,p) = 
16 \pi \; (qp)^{J} \; a_{4\pi}^2 
\; e^{-q^2/\Lambda_{4\pi}^2} \; e^{-p^2 / \Lambda_{4\pi}^2},
\label{V_4Pi}
\eeq
and the four-kaon term $K^{(4K)J}(q,p)=0$.
The second four-point interaction is a short-ranged attraction modeled 
as a $t$-channel exchange of a heavy scalar-isoscalar meson.  
Its form is given by the scalar-exchange kernel of \Eq{ExplicitK1J} and
the two-body self-energy $\Sigma$ of \Eq{Sigma1Int}.
For simplicity it is treated exactly as if two additional three-body
states, 
\bea
\ket{\pi\pi X}, \ket{ K\bar{K} X},
\eea 
with $m_{X}=$ 1500~MeV, were added to the Hilbert space.
Clearly, modelled in this manner, for energies $E > m_{X}+2m_{\pi}$ =
1780~MeV, the state $\ket{\pi\pi X}$ can go on-energy-shell.
However, the calculations described herein are for energies less than
1400~MeV, so that $\ket{\pi\pi X}$ is never on-energy-shell.

Again, the objective is to study the framework developed in this paper.  
That is, it is interesting to assess the importance and possibility of
including the dynamics of three-body intermediate states to the study of
meson scattering and final state interactions, rather than to test a
particular interaction model for $\pi\pi$ scattering. 

% ********************************************    coupling constants.
 \begin{table}
 \caption{Coupling strengths and momentum scales for the vertex form
 factors.  The values of the momentum scales are given in MeV.}
 \begin{tabular}{cccccc}
 $\beta_1\beta_2\alpha_1$ & $\pi\pi f_0$& $K\bar{K}f_0$ 
        &$\pi\pi\rho$&$K \bar{K}\rho$&$K\pi K^*$ \\ \hline
$a_{\beta_1\beta_2\alpha_1}$ & 12.4 & 5.08 & 20.0 & 20.0 & 20.0  \\
$\Lambda_{\beta_1\beta_2\alpha_1}$& 875 & 875 & 875 & 875 & 875 \\
 \end{tabular}
 \label{Tab:Couplings}
 \end{table}
%********************************************************************
Once the forms of the model vertices are fixed, the only observables used 
in the fit are the
scalar-isoscalar $\pi\pi$ phase shift $\delta_{\pi\pi}$, the
existence of the $K\bar{K}$ resonance (referred to as the $f_0(980)$),  
and the decay width of the $\rho(770)$ vector meson.  
All measurable reaction channels are not used in the fitting procedure,
since this paper represents more a proof of principle of the framework
rather than a complete phenomenological analysis of $\pi\pi$ scattering.   
The resulting values of the coupling constants are provided in
Tables~\ref{Tab:Couplings} and \ref{Tab:Couplings2}.   

% ********************************************    coupling constants.
 \begin{table}
 \caption{Coupling strengths and momentum scales for the four-point meson
 interactions. The values of the momentum scales are given in MeV. }
 \begin{tabular}{ccccc}
        & $\pi\pi X$& $K\bar{K} X$ 
        & $4\pi$& $4K$ \\ \hline
$a$ & 19.8 & 11.0 & 17.2 & 0.0 \\
$\Lambda$& 1000 & 1000& 750 & 750 \\
 \end{tabular}
 \label{Tab:Couplings2}
 \end{table}
%********************************************************************
%.......................................................
\subsection{Phase shifts and inelasticities}
\label{Sec:PhiS}
%
%*********************************************** Plot of pipi phase shift.
\begin{figure}[tb]
\epsfig{figure=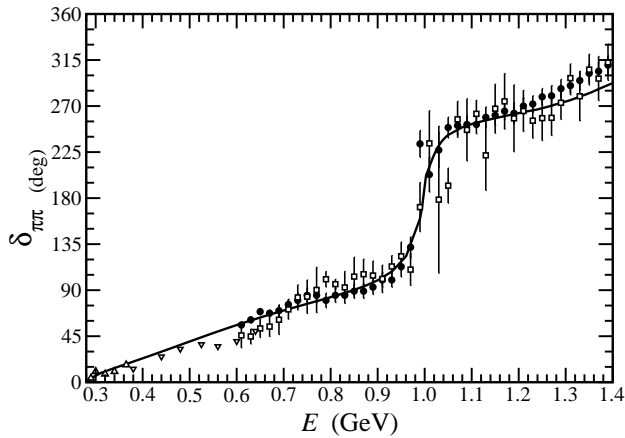,width=7.0cm,angle=-90}
\caption{The $\pi\pi$ scattering phase $\delta_{\pi\pi}$ in the scalar,
isoscalar channel as a function of the CM driving energy $E$.  
The data are from Refs.~\protect\cite{Kaminski99} (open squares), 
\protect\cite{CernMunich} (closed circles),  \protect\cite{PiPiData1} 
(up triangles), and \protect\cite{PiPiData2} (down triangles).} 
\label{Fig:PhiS}
\end{figure}
%*********************************************** Plot of inelasticity.
\begin{figure}[tb]
\epsfig{figure=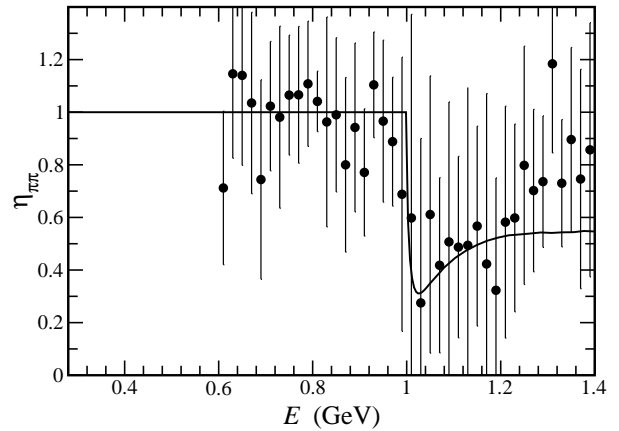,width=7.0cm,angle=-90}
\caption{The $\pi\pi$ scattering inelasticity $\eta_{\pi\pi}$
in the scalar, isoscalar channel as a function of the CM driving energy
$E$.  The data are from Ref.~\protect\cite{Kaminski99}.} 
\label{Fig:EtaS}
\end{figure}
%*******************************
Below all three-body thresholds, the non-trivial part of the $S$-matrix in
\Eq{Def:TMatrix} can be written as a $2\times 2$ unitary matix
$S_{\beta\alpha}(E)$ with $\alpha,\beta$ denoting the only two open
channels $\pi\pi$ and $K\bar{K}$.   
The $S$-matrix in the $J^{\rm th}$ partial wave can be parametrized in
terms of two phase shifts $\delta_{\pi\pi}$ and $\delta_{K\bar{K}}$, and
one inelasticity $\eta_{\pi\pi}$, 
\bea
\delta_{\alpha}(E) &=&   \frac{-i}{2 } \ln
        \frac{S^{J}_{\alpha\alpha}(E)}{ \eta_{\alpha}(E)} ,  
        \label{Def:Phi} \\
\eta_{\alpha}(E) &=&  |S^{J}_{\alpha\alpha}(E)| ,
        \label{Def:Eta}
\eea
for $\alpha = {\pi\pi}, {K\bar{K}}$.  
Below all three-body thresholds there are only two stable channels, 
${\pi\pi}$ and ${K\bar{K}}$.  Hence, there is only one inelasticity
parameter $\eta_{\pi\pi}=\eta_{K\bar{K}}$.   
For energies $E$ above the lowest stable three-body threshold,
one must augment the $S$-matrix by including all 
stable three-body states.  Consequently, its parametrization requires more
than two phase shifts and one inelasticity.
Nonetheless,  one may still use \Eqs{Def:Phi} and \rf{Def:Eta} to define
the phase shifts $\delta_{\alpha}(E)$, and inelasticities
$\eta_{\alpha}(E)$ for the  two channels $\alpha=\pi\pi$ and $K\bar{K}$.
Of course, above the threshold of a stable three-body state
$\eta_{\pi\pi}\not=\eta_{K\bar{K}}$. 

% Initial discussion of our results.
The $\pi\pi$ phase shift, as defined by \Eq{Def:Phi} and obtained from our
model, is shown as a solid curve in \Fig{Fig:PhiS}.    
The model provides an excellent description of the $\pi\pi$ phase shift
data depicted in \Fig{Fig:PhiS} 
\cite{Kaminski99,CernMunich,PiPiData1,PiPiData2}, and the 
inelasticity $\eta_{\pi\pi}$, as shown in \Fig{Fig:EtaS}. 
The overall trend of the pion scattering phase shift $\delta_{\pi\pi}$ is
positive and increases slowly with energy $E$.  This is indicative of a weak
and attractive effective $\pi\pi$ scattering potential.  At the kaon
threshold, a rapid phase motion is apparent.  It results from the presence
of a narrow, $f_0(980)$ scalar meson.  
Above the two-kaon threshold, the phase shift continues to increase slowly, 
at a rate similar to the increase in the phase shift below the threshold.

Below all other thresholds, the $\pi\pi$ channel is the only open channel,
and unitarity requires that the inelasticity $\eta_{\pi\pi}=1$ here.
This is clearly observed in \Fig{Fig:EtaS}, where the calculated
inelasticity $\eta_{\pi\pi}$ has a value consistent with unity below the
threshold of the $K\bar{K}$ channel at 1~GeV.
Had four-body states been included into this framework, one might have
expected to see a decrease in the inelasticity $\eta_{\pi\pi}$ due to the
opening of the four-pion state, which has a threshold of
$E=4m_{\pi}\approx$ 0.560~GeV.  However, the data in \Fig{Fig:EtaS} 
\cite{Kaminski99} seem to suggest that the contribution
of the four-pion state to $\pi\pi$ scattering is negligible.
This can be seen by noting the lack of any systematic deviation from
$\eta_{\pi\pi}=1$ for the range of energies $4m_{\pi} < E < 2 m_{K}$. 

%***************  Importance of KKbar discussion ********************
It is clear from both \Figs{Fig:PhiS} and \ref{Fig:EtaS} that the
$K\bar{K}$ channel has a significant effect on $\pi\pi$ scattering.   
At the two-kaon threshold at $E\approx$ 1.0~GeV,  
one observes a rapid increase in the $\pi\pi$ phase shift
$\delta_{\pi\pi}$, and a sharp fall off of the $\pi\pi$ inelasticity
$\eta_{\pi\pi}$ to its minimum value $\eta_{\pi\pi}\approx$ 0.31. 
%....................
\begin{figure}[tb]
\epsfig{figure=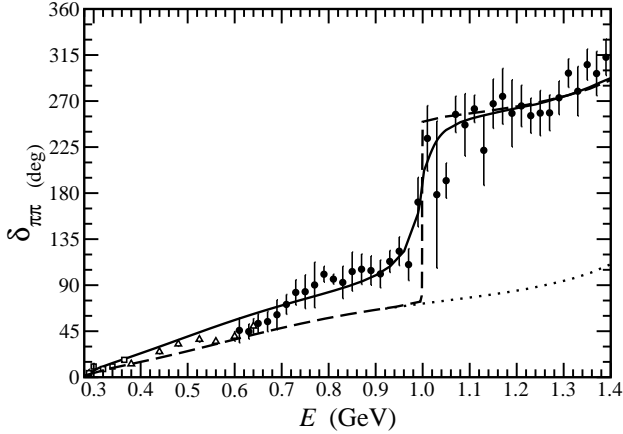,width=7.0cm,angle=-90}
\caption{The $\pi\pi$ scattering phase $\delta_{\pi\pi}$ in the scalar,
isoscalar channel versus the CM energy $E$.  
The full calculation (solid curve) is compared to the calculation (dashed
and dotted curves) in the limit the ${K\bar{K}}$ and $\pi\pi$ states
become decoupled.  The data are from Ref.~\protect\cite{Kaminski99}.}  
\label{Fig:PhiSNoK}
\end{figure}
%........................
This phase motion is indicative of crossing the thresholds of the two-body
$K\bar{K}$ state and the one-body $f_0(980)$ bound state.  
The rapid increase observed in the phase shift $\delta_{\pi\pi}(E)$ is
due to the {\em weak} coupling between the $\pi\pi$ and $K\bar{K}$
channels. 
In the model, when the mixing of these two channels is further weakened,
the rate of change of the phase motion tends to {\em increase},  
until finally, in the limit that the coupling between the two channels 
goes to zero, the phase motion becomes a step-function of magnitude 
180 degrees.  Such a phase motion is completely unobservable, and could
therefore be ignored altogether (although it is relevant to Levinson's
theorem, which relates overall changes in the phase shifts from 
threshold to infinite energy to the number of bound states in 
a system).  

The importance of the coupling between the two-body channels $\pi\pi$ and
$K\bar{K}$ can be estimated quantitatively by recalculating the pion phase
shift $\delta_{\pi\pi}$ after removing the kaon state $\ket{K \bar{K}}$ from
the Hilbert space.  
The result is shown as the dotted curve in \Fig{Fig:PhiSNoK}.  
However, from the above argument, perhaps a better indication of the
importance of the $K\bar{K}$ channel is obtained by letting the 
couplings that lead to a mixing between the $K\bar{K}$ and $\pi\pi$
channels go {\em smoothly} to zero. 
(In practice, this is done by not allowing the one-body scalar $f_0(1350)$
to have a bare coupling to the $K\bar{K}$ state, and setting $a_{K^*\pi K}=0$.
This has a minimal effect on the dynamics but prevents mixing the
$\ket{\pi\pi}$ and $\ket{K\bar{K}}$ states.)
In this limiting case, the two states $\ket{K\bar{K}}$ and
$\ket{f_0(980)}$ are coupled to the $\ket{\pi\pi}$ state, but the
contributions they make to $\pi\pi$ scattering go to zero.
The result is that for energies $E >$ 1000~MeV, the $\pi\pi$ scattering
amplitude will cross the branch cuts associated with the
opening of these two channels.
But since these channels do not mix with the $\pi\pi$ channel in this
limit, the $\pi\pi$ phase shift exhibits a step-like motion at $E=$ 1.0~GeV.
This is shown as the dashed curve in \Fig{Fig:PhiSNoK}.
The difference between the solid curve, which represents the full model
calculation, and the dashed curve in \Fig{Fig:PhiSNoK} may be taken as
being indicative of the significance of the $K\bar{K}$ channel on $\pi\pi$
scattering.  
One concludes that the mixing between the $\pi\pi$ and $K\bar{K}$ channels
is significant near and below the two-kaon threshold, where it can
contribute more than half of the total phase shift $\delta_{\pi\pi}$.
At energies above the two-kaon threshold, its importance quickly dimishes
and vanishes altogether above $E=$ 1150~MeV.

%*********************************************** Plot of eta with NO KAONS.
 \begin{figure}
 \epsfig{figure=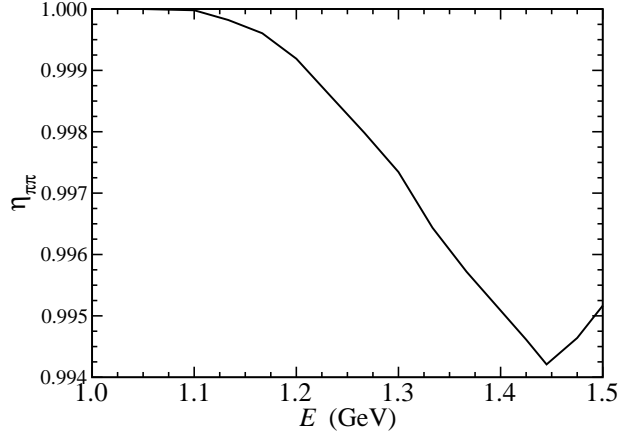,width=7.0cm,angle=-90}
 \caption{The $\pi\pi$ inelasticity $\eta_{\pi\pi}$ in the
 isoscalar-scalar channel with no kaons.}
 \label{Fig:EtaNoK}
 \end{figure}
%*****************************
The importance of the mixing between the $\pi\pi$ and $K\bar{K}$ channels
can also be observed in the pion inelasticity $\eta_{\pi\pi}$ shown in
\Fig{Fig:EtaS}.   Just above the two kaon threshold, the inelasticity
plummets to a minimum value $\eta_{\pi\pi} \approx 0.31$.   
In the limit that the coupling to the kaon channel goes to zero, as
described above, the inelasticity takes on a very different appearance.

In \Fig{Fig:EtaNoK}, the two pion inelasticity is plotted above the
$K\bar{K}$ threshold.  Below the threshold its value is unity.
The lowest multi-particle state to which two-pion flux can be lost is the
three-particle channel $\ket{\pi\pi\rho}$.  The production threshold of 
this state is 1050~MeV.  
It is clear that one observes a slow decrease in the inelasticity
$\eta_{\pi\pi}$ above 1050~MeV, due to the opening of the $\pi\pi\rho$
channel. The effect of this channel on the inelasticity is very small, 
with a minimum value $\eta_{\pi\pi}\approx 0.994$. 

%*********************************************** Argand plot.
\begin{figure}[tb]
\epsfig{figure=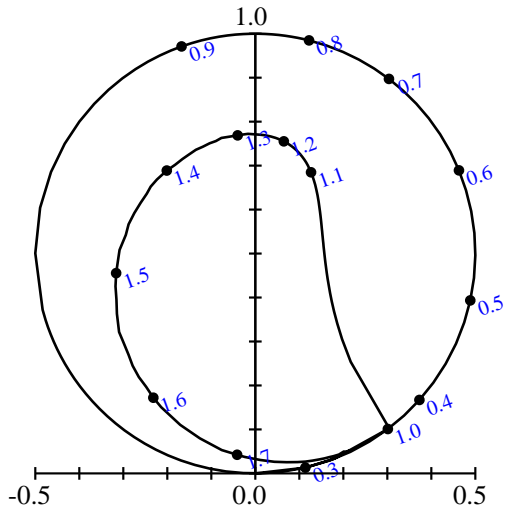,width=7.8cm,angle=-90}
\caption{Argand diagram for $\pi\pi$ scattering at all energies.
        Plotted is the energy dependence of the real and imaginary parts
        of the partial-wave scattering amplitude $a_{\pi\pi}^{J=0}(E)$ for
        the $J=0$ partial wave from \protect\Eq{Def:a1} or
        \protect\Eq{Def:a2}. 
        The curve is calculated for {\em all} energies $E>2m_{\pi}$,  
        and annotated with the corresponding energies $E$ in GeV, from
        threshold to above $E=$ 1.7~GeV. 
        }
\label{Fig:Argand}
\end{figure}
%************************************************************
The energy dependence of the $S$-wave pion scattering inelasticity
$\eta_{\pi\pi}$ and phase shift $\delta_{\pi\pi}$ are conveniently
plotted together in an Argand diagram.
In \Fig{Fig:Argand}, the partial-wave scattering amplitude 
$a^{J=0}_{\pi\pi}(E)$ is plotted in the complex plane as a parametric 
function of the CM energy $E$. This figure clearly shows the 
rapid rise in the elastic $\pi\pi$ phase shift just below 1 
GeV, and the resulting dramatic loss of flux from the elastic 
channel as soon as the $K\bar{K}$ channel opens up.   

\begin{figure}[tb]
\epsfig{figure=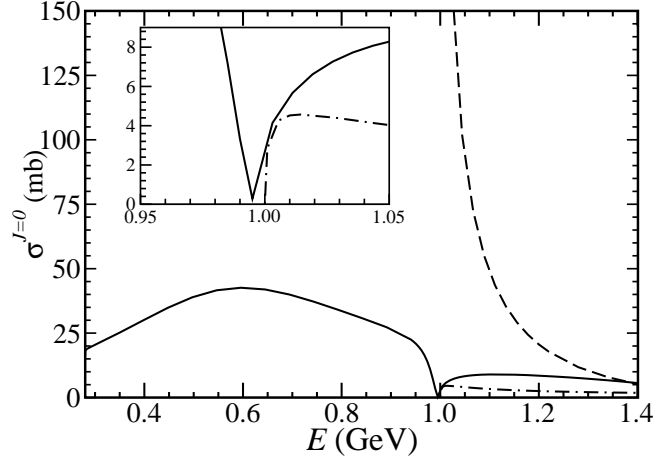,width=7.35cm,angle=-90}
\caption{$S$-wave cross section $\sigma^{J=0}(E)$ in millibarns for
$\pi\pi\rightarrow\pi\pi$ (solid curve), 
$K\bar{K}\rightarrow K\bar{K}$ (dashed curve), and
$\pi\pi\rightarrow K\bar{K}$ (dot-dashed curve).
The inset shows a detail of the region around the $K\bar{K}$ threshold at
$E=$ 1~GeV.} 
\label{Fig:XSection}
\end{figure}

For spinless particles, the amplitude is given by 
\bea
a^{J}_{\pi\pi}(E)&=& \frac{\eta_{\pi\pi} e^{2i\delta_{\pi\pi}} - 1}{2i}, 
        \label{Def:a1}
\\
&=& - \frac{p_0}{(4\pi)^2} \frac{1}{2E} z_{\pi\pi} \nonumber \\
&& \times
        \left( t^{(1)J}_{\pi\pi,\pi\pi}(p_0,p_0) + 
                t^{(2)J}_{\pi\pi,\pi\pi}(p_0,p_0) 
        \right),        \label{Def:a2}
\eea
where $p_0=\sqrt{E^2/4 - m_{\pi\pi}^2}$ is the magnitude of the
on-energy-shell three-momentum of the pions, 
$z_{\pi\pi}$ is the wave function renormalization of the two-pion state
$\ket{\pi\pi}$, 
and $t^{(1)J}_{\beta\alpha}(q,p)$ and $t^{(2)J}_{\beta\alpha}(q,p)$
are the two-body scattering amplitudes, obtained from 
\Eqs{eq:t22sol} and \rf{t1fin}, respectively. 
The two-body scattering cross section $\sigma_{\beta\leftarrow\alpha}(E)$
can be written in terms of the partial-wave scattering cross sections
according to 
\bea
\sigma_{\beta\leftarrow\alpha}(E) = 
        \sum_{J=0}^{\infty} \sigma^{J}_{\beta\leftarrow\alpha}(E).
\eea
In terms of the scattering amplitudes, these partial-wave cross sections
are given by 
\bea
\lefteqn{
\sigma_{\beta\leftarrow\alpha}^{J}(E)=
}\nonumber \\
&& 
        \frac{2J+1}{4\pi}
        \frac{z_{\beta} z_{\alpha}}{64\pi^2 E^2} 
        \left| 
        t^{(1)J}_{\beta\alpha}(q_0,p_0) + 
                t^{(2)J}_{\beta\alpha}(q_0,p_0) 
        \right|^2 ,
\eea
where $p_0$ and $q_0$ are the on-energy-shell solutions to 
$\M_{\alpha}(p_0)=E$ and $\M_{\beta}(q_0)=E$, respectively.
The resulting cross section for elastic, $S$-wave $\pi\pi$ scattering is
shown as a solid curve in \Fig{Fig:XSection}.
It is finite at threshold, exhibits a maximum value of 43 millibarns at
$E\approx$ 600~MeV, and a sharp decrease at the position of the
$f_0(980)$ scalar meson resonance.  
This sudden drop occurs just below the $K\bar{K}$ threshold, as can be seen
upon examination of the inset plot in \Fig{Fig:XSection} which depicts a
closeup of the $K\bar{K}$ threshold region.

The dot-dashed curve is the resulting cross section
$\sigma^{J=0}_{K\bar{K}\leftarrow\pi\pi}$ for the two-kaon production
process $\pi\pi\rightarrow{K}\bar{K}$.  
This cross section is considerably smaller than that of the elastic
$\pi\pi$ scattering cross section, reaching its maximum value of 4.6
millibarns just above the two-kaon threshold at $E=$ 1~GeV.
Its small size is a result of the weak coupling between the two-pion and
two-kaon channels.  
As discussed above, a weak coupling of these channels is necessary to
ensure a narrow $f_0(980)$ meson.  
If the mixing between the pion and kaon channels were stronger, the
$f_0(980)$ meson would more easily decay into two pions, tending 
to increase its width significantly.

The dashed curve in \Fig{Fig:XSection} is the $S$-wave cross section for
elastic $K\bar{K}$ scattering.   This cross section is comparably huge,
having its maximum at the two-kaon threshold energy.   Its size can be
compared to that of the two-pion elastic cross section,
\bea
\sigma_{\pi\pi\leftarrow{\pi\pi}}^{J=0}(E=2m_{\pi})
        &\approx& 18.3~{\rm mb}, \nonumber \\
\sigma_{K\bar{K}\leftarrow{K}\bar{K}}^{J=0}(E=2m_{K})
        &\approx& 734~{\rm mb}. \nonumber 
\eea
The very large $K\bar{K}$ cross section arises from the scalar 
$f_0(980)$ meson which lies just below the $K\bar{K}$.  
The presence of a bound state just below the  two-body scattering threshold
will generally tend to increase the size of the cross section
dramatically.  

%**************************************** Cross sections.
%*******************************

%***************  Pion phase shift DATA ********************
As was discussed earlier in \Sec{Sec:Model}, the model parameters were chosen
to provide a good fit to the $S$-wave pion phase shift $\delta_{\pi\pi}$
data from \Ref{PiData}.  In \Fig{Fig:PhiS}, the resulting phase shifts are
also in excellent agreement with the data analyses of 
\Refs{Kaminski99}, \cite{PiPiData1}, and \cite{PiPiData2}.
However, it is important to realize that direct comparison of these 
model results with these $\pi\pi$ scattering data must be done with
caution.  
Extraction of these data requires the use of theoretical models, or
theoretical assumptions, in order to fit the experimental observables.
In the worst-case scenario, the resulting $\pi\pi$ phase shift data may be
more representative of the extraction methods employed than of the actual
$\pi\pi$ scattering process.   

The extraction of the $\pi\pi$ phase shifts from experiment is a
difficult and long-standing problem of hadron physics.  
At present, it is impossible to construct an experiment in which a pion
beam is scattered from a pion target.  Hence, other techniques are
required to extract the $\pi\pi$ phase shift from experimental observables.
One possibility is to use decays that produce two pions in the final
state, and attempt to extract the $\pi\pi$ phase shifts from the
final-state interactions.  

\begin{figure}[tb]
\epsfig{figure=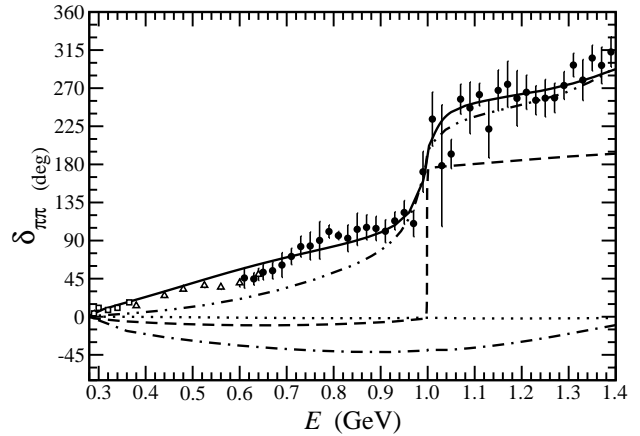,width=7.0cm,angle=-90}
\caption{Importance of the different model contributions to $\pi\pi$
scattering phase shift.  The solid curve is the full calculation.  
The dashed curve is obtained by removing the one-body state $f_0(1350)$
from the Hilbert space.  
The dot-dashed curve is obtained by completely removing the scalar 
$f_0(1350)$ from the theory, both in one-body and three-body states.
The dotted curve is same as the dot-dashed curve, but all scalar couplings
(the ``$X$''-exchanges and contact terms) are also set to zero.  
This is the effect of just the vector-meson exchanges.
Above threshold it is negative with a minimum of
$\delta_{\pi\pi}\approx$ -2.0 degrees at $E\approx$ 1.2~GeV.
The dot-dot-dashed curve is obtained by weakening the coupling of the 
one-body $f_0(1350)$ state to the $\pi\pi$ state by 10 percent. The data
points are from \protect\Fig{Fig:PhiS}.}
\label{Fig:PhiS_Contrib}
\end{figure}

The procedure employed by \Ref{PiPiData1} is extract the phase shifts
from the electroweak kaon decay $K^+\rightarrow\pi^+\pi^0e^+\nu_e$.
The results are shown as open squares in \Figs{Fig:PhiSNoK} and 
\ref{Fig:PhiS_Contrib}.
Extraction of the phase shift using decays with more than two particles in
the final state requires some knowledge of transition form factors for
the coupling of a kaon, two pions and the $W$ boson to determine
$K^{+}\rightarrow\pi^{+}\pi^{-}W^{+}$.  
A nice feature of employing the electroweak decay
$K^+\rightarrow\pi^+\pi^0e^+\nu_e$ is that the two pions are the only
strongly-interacting particles in the final state.
Hence, one would expect that the $\pi\pi$ interactions would be the
dominant contribution to the dressing of the final state. 
However, this approach is hampered by two experimental difficulties. 
The first is the lack of statistics.  This particular kaon 
decay represents only a small fraction ($4 \times 10^{-5}$) of the 
total $K^+$-meson decay width, which is already extremely small.   
The second is the fact that the energies and angles of the outgoing leptons 
provide a small lever arm with which to vary the CM energy $E$ of the two
final-state pions.

Another method is to extract the final state interactions of the two-pion
production process ${\pi}p\rightarrow\pi\pi{n}$, employed by
\Ref{PiPiData2} (down triangles), \Ref{Kaminski99} (open squares),
and \Ref{CernMunich} (closed circles), shown in \Fig{Fig:PhiS}. 
These studies require some theoretical input in order to perform
the extraction of the $\pi\pi$ scattering phase shift $\delta_{\pi\pi}$
and inelasticity $\eta_{\pi\pi}$; hence, they are not direct measurements
of the $\pi\pi$ scattering phase shift.  

Our model parameters were originally fit to the $\pi\pi$ phase shifts
obtained by an analysis~\cite{CernMunich} of an experiment at CERN
involving $\pi^{-}p\rightarrow\pi^+\pi^{-}n$ at 17.2~GeV.
Recently, this same data was re-examined by Kaminski {\it et al.} 
\cite{Kaminski99}, with weaker
model assumptions than were used in \Ref{CernMunich}.
The work of \Ref{Kaminski99} provides an exhaustive and {\em nearly}
complete study of the $\pi\pi$ phase shift.
In particular, there is no assumption that pion exchange is the dominant
mechanism for the process $\pi^{-}p\rightarrow\pi^+\pi^-n$.  
Consequently, this analysis seems to be more general than the others.
A relative phase ambiguity in the analysis of \Ref{Kaminski99} provides
four possible, distinct solutions for the $\pi\pi$ phase shifts and
inelasticities.   Two of these solutions seem to have an unphysical
inelasticity $\eta_{\pi\pi}$ below the two-kaon threshold and can be
discarded.   
The other two solutions, denoted the ``up-flat'' and ``down-flat''
solutions, are very similar in appearance and neither can be dismissed on
qualitative grounds.   The phase shifts from the ``down-flat'' solution of
\Ref{Kaminski99} are shown as solid circles in \Figs{Fig:PhiSNoK} and
\ref{Fig:PhiS_Contrib}. 

The general behavior of the $\pi\pi$ phase shift $\delta_{\pi\pi}$ is
positive, which is indicative of an attractive $\pi\pi$ scattering
amplitude.
An abrupt increase in the pion phase shift is evident at 1~GeV, which is
due to the combined effects of the opening of the $K\bar{K}$ threshold and
the crossing of the scalar $f_0(980)$ resonance.   

Apart from this feature, which in this particular model is a result of the
delicate mixing between the $\pi\pi$ and $K\bar{K}$ channels, 
the calculated pion phase shift $\delta_{\pi\pi}$ exhibits a steady,
gentle increase from the threshold at 280~MeV to above 1400~MeV!  
In the model, this slowly increasing behavior arises from subtle 
cancellations between the attractive potentials of the heavy scalar-meson
exchanges and the repulsive, direct four-pion interaction from \Eq{V_4Pi}
which is intended to model the effect of dynamical chiral symmetry
breaking. 
The different form factor scales involved in these interactions (see
Tables~\ref{Tab:Couplings} and \ref{Tab:Couplings2}) are chosen to provide
this slowly increasing, weak phase shift observed in the two-pion channel.
Typically,  scalar potentials by themselves provide a strong attraction in
the $S$-wave $\pi\pi$ channel, that leads to a rapidly rising phase shift
just above the $\pi\pi$ threshold, which then quickly falls away.
This behavior is not seen in the $\pi\pi$ phase shift $\delta_{\pi\pi}$.

%...............................
%...............................
The importance and role of the scalar resonances in $\pi\pi$ scattering can
be appreciated by a close examination of \Fig{Fig:PhiS_Contrib}.  
The effect of the heavy $s$-channel resonances 
which are collectively modeled by the single $f_0(1350)$ in the model is
two fold. First, their presence leads to a strong attractive potential for
energies below 1400~MeV.  Second, they provide the most important
contribution to {\em off-diagonal} matrix elements of the two-body scattering
kernel $K$.  That is, they provide the strongest source of mixing for the
two-pion and two-kaon states in this model.

Both of these effects tend to produce an attraction for the two pions. 
In particular, the strong attraction necessary to bind the kaons to form
the $f_0(980)$ resonance results in a strong attraction in the two-pion
channel as well.  
The amount of mixing between the two-pion and two-kaon states dictates the
attraction felt by the pions.   
Hence, when the coupling between $\ket{\pi\pi}$ and $\ket{f_0(1350)}$
states is artificially reduced by as little as 10\%, the result is
significant, as shown by the dot-dot-dashed curve in \Fig{Fig:PhiS_Contrib}.  
When the one-body state $\ket{f_0(1350)}$ is removed entirely from the
Hilbert space, the result is the dashed curve.  
The resulting pion phase shift is negative and close to zero below the
two-kaon threshold, and positive above the threshold.  
The absence of the one-body $s$-channel state $f_0(1350)$ reduces the
mixing between the two-pion and two-kaon states, which results in a nearly
stable (and very narrow) kaon bound state $f_0(980)$.   
In this case, one observes a $f_0(980)$ bound state with a width that has
been reduced from 46~MeV to 0.28~MeV!   
This is a result of the fact that the $f_0\rightarrow\pi\pi$ decay must
proceed through $K^*$ exchange in the kernel $K$, which provides only a
weak mixing of the $\pi\pi$ and $K\bar{K}$ states.

When all scalar mesons are removed from the theory entirely; that is, when
the couplings that lead to the existence of one-body states
$\ket{f_0(1350)}$,  and three-body states $\ket{f_0\pi\pi}$ and
$\ket{f_0K\bar{K}}$ are set to zero $a_{\pi\pi f_0}=a_{K\bar{K}f_0}=0$, 
the resulting phase shift is shown in \Fig{Fig:PhiS_Contrib} by the
dot-dashed curve.  The slightly repulsive behavior is a result of the
combined effect of the attractive scalar-$X$ and repulsive four-point
interactions associated with chiral symmetry, given by the parameters in
Table~\ref{Tab:Couplings2}. 

To minimize the number of free parameters all of the couplings to vector
mesons are chosen to be equal to each other 
$a_{\pi\pi\rho}=a_{K\bar{K}\rho}=a_{\pi K K^*}$.
These coupling strengths were then determined by solving $P$-wave $\pi\pi$
scattering at the $\rho$ meson mass $E=m_{\rho}=$ 770~MeV, and requiring
that the $\rho$-meson width reproduced the experimental value 
$\Gamma_{\rho}=$ 150~MeV, as given in Table~\ref{Tab:Particles}.
It is found that the resulting coupling strength leads to a vector-meson
exchange interaction kernel $K$ which provides a very weak repulsion for
$\pi\pi$ scattering. 
This is illustrated by the dotted curve in \Fig{Fig:PhiS_Contrib}, where
all of the couplings except those involving the vector 
mesons ($a_{\rho\pi\pi}$, $a_{\rho K {\bar{K}}}$, $a_{K^* \pi \bar{K}}$)
are set to zero.
The resulting phase shift is negative (repulsive) and very small; its
largest absolute value is about 2 degrees.   
Thus, in this model, $\rho$-meson exchange is negligible in the $S$ wave.
This model differs from the analysis of \Ref{Speth95}, in which they
report that the attractive potential (which leads to the binding of the
$K\bar{K}$ into the $f_0(980)$ resonance) is primarily due to   
$\rho$-meson exchange, which is strong and attractive in their model.  
In the framework, the exchange of a spin-1 meson in the kernel $K$ results
in a very weak and mildly repulsive interaction.  
This difference is a result of how the spin couplings and energy
denominators of the meson-exchange propagators are implemented in the
kernels of the two frameworks.

%.......... Brief discussion of unitarity.

Before ending this section, a final comment concerning the accuracy of the
numerical methods employed is provided.   The accuracy is measured using
the unitarity condition (or optical theorem), 
\beq
T - T^{\dagger} = T^{\dagger} ( G - G^{\dagger} ) T \label{UnitAbstract},
\eeq
which is derived from the fact that $V$ in \Eq{VMatrix} is Hermitian. 
Evaluating \Eq{UnitAbstract} between two-pion states, one obtains an
equation that relates the two-pion flux missing from the {\em forward}
direction to the one-, two-, and three-body outgoing flux observed
leaving the scattering center.  This relation provides a sensitive check
of the numerical methods employed.   

\begin{figure}
\epsfig{figure=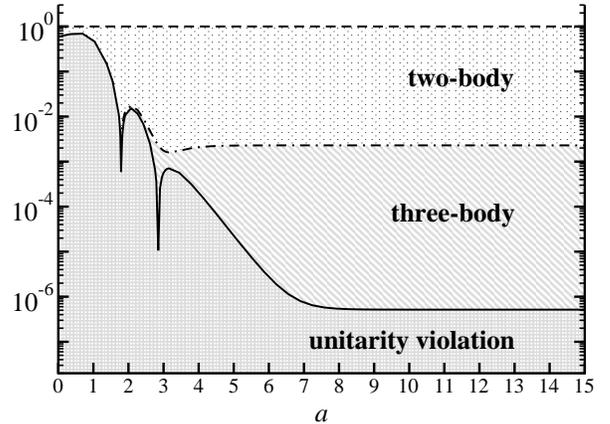,width=7.0cm,angle=-90}
\caption{
Fraction of incoming two-pion flux at energy $E=$ 1400~MeV
in the isoscalar-scalar channel, appearing as ``two-body'' or 
``three-body'' final states, or lost to unitarity violations.
The fractions are shown versus the exponent $a$ of the adiabatic scale
$\epsilon=10^{-a}$~GeV.  
}
\label{figunit}
\end{figure}

The fraction of lost two-pion flux observed as outgoing two- or
three-particle states is shown versus the exponent $a$ of the 
adiabatic scale $\epsilon=10^{-a}$~GeV in \Fig{figunit} for CM energy  
$E=1400$~GeV. 
For any energy $E$ greater than the two-pion scattering threshold, 
there can be no stable one-particle state.  
It follows that no one-particle outgoing flux can be observed.   
In \Fig{figunit}, the regions labeled ``two-body'' and ``three-body''
represent the fractions of the missing two-pion flux that
appear as outgoing two-body ($\pi\pi$ or $K\bar{K}$) and three-body
($\pi\pi\rho$) states, respectively.   
Since the Hilbert space is restricted to one-, two- and three-body states,
this should account for all the flux.  However, in practice, the numerical
methods employed introduce violations to the unitarity condition of
\Eq{UnitAbstract}.  
The region below the solid curve in \Fig{figunit} represents the fraction
of flux that completely disappears from the theory; 
such a loss of flux violates unitarity.  
One observes that for values of $\epsilon \le 10^{-7}$ GeV, the violation
is less than $10^{-6}$ of the outgoing flux and is therefore negligible.  
In the application to $\pi\pi$ scattering described above, 
the value of $\epsilon=10^{-12}$~GeV (or $a=12$) was employed.

%...................................................................
%...................................................................

%***********************************************************************
\section{Summary and future directions}
\label{Sec:Summary}
Herein, a framework suitable for the description of nonperturbative hadron
scattering, based on the Bakamjian-Thomas formulation of relativistic
quantum mechanics is introduced.    
In \Sec{Sec:Formalism}, it is shown that by including the interactions
into the free mass operator $\M$, one can ensure that observables
calculated in the framework are Lorentz coviariant.   
When the Hilbert space is truncated to contain only one-, two- and
three-body states, the resulting Lippmann-Schwinger equations (LSEs) form
a closed set of coupled integral equations.
The solution of these integral equations is obtained numerically.

A significant improvement of this framework over earlier work 
is that the {\em full} effect of three-body states have been included.
The three-body Green functions appear only in the two-body scattering
kernel $K$ and the two-body self-energy $\Sigma$, and for energies $E$
above the three-body thresholds, unitary branch cuts associated with these
thresholds appear both in $K$ and $\Sigma$.
It follows that the kernel $K$ and two-body self-energy $\Sigma$ are 
{\em complex} functions of the energy $E$.  
The appearance of such three-body branch cuts provides the means for
important dynamical effects, such as three-hadron production and decays
into three-hadron states, which are automatically accounted for in our
framework.   Such effects have hitherto been ignored for the most part in
previous studies.

The inclusion of three-body cuts into the integral equations requires 
appropriate numerical methods to be employed.   
These methods, necessary to solve the coupled set of LSEs are shown to
maintain the unitarity of the theory to better than one part in one
million.  
To demonstrate the utility of the framework, a preliminary study
of $\pi\pi$ scattering is carried out in \Sec{Sec:PiPi}.
The simple model, introduced in \Sec{Sec:Model}, is able to provide an
excellent description of the $\pi\pi$ phase shifts and inelasticities. 

The main purpose for developing this framework is to provide a means of
incorporating the dynamics of low-momentum transfer, final-state
interactions into the study of hadronic processes for energies up to a
couple of GeV. 
In this energy region, the comparison between experimental data and 
theorectical predictions from models of QCD for quark and gluon
dynamics, are often made difficult due to the presence of final-state
interactions.  
The soft rescattering of final-state hadrons tends to {\em mask} the
QCD dynamics of interest.  
The framework developed herein provides a tractable means to incorporate
the effects of final-state interactions into studies of hadronic
phenomena. 

Towards this end, the framework is constructed to be a consistent
extension for the constituent quark model. It provides a means to 
{\em unquench} the quark model by providing for hadron loops,
multi-particle thresholds and the unitarity branch cuts associated with
these.  
The result is the generation of complex-valued scattering amplitudes.
It is an extension of the quark model, in that the quark model may be used
to provide the elementary couplings and form factors for the hadronic 
interactions in $V$.  The framework uses this real potential $V$ to
generate the full scattering solution.

Future applications of the framework will focus on the dynamics of 
nucleon resonances and exotic, hybrid mesons that are the subject of current 
 and proposed experiments at TJNAF. 
In the baryon case, for energies up to about 2~GeV, there are some well-known
and striking examples in $\pi N$ scattering for which three-body effects
are crucial in understanding the experimental observables.
For example, in the $L_{2I,2J}=P_{11}$ channel, the $\pi N$ inelasticity
arising from the three-body $\pi\pi N$ state is very large~\cite{Lee86}.  
It is likely that a complete understanding of the $P_{11}$ $\pi N$
scattering channel and the mysterious $N^{*}(1440)$ resonance requires the
full implementation of three-body unitary cuts that this framework
provides.  

The effect of three-body cuts in exotic partial waves can also be 
very important. Most theoretical studies of hybrid meson decays have so 
far ignored effects of final state interactions. The best candidate for the 
exotic meson, the $\pi_1(1600)$~\cite{Adams}  
was found in the $\rho\pi$ decay channel, which is predicted to be suppressed 
with respect to other two body decay channels, in particular the 
$b_1\pi$~\cite{Isgur,PPASES}. 
This shift of strength from $b_1\pi$ to $\rho\pi$ could be explained by mixing 
with the three-body, $\omega\pi\pi$ intermediate state which is believed
to have a strong coupling to these two--meson channels.

%***********************************************************************
%.......................................................
%
% Acknowledgments.
%
%.......................................................
% \vspace*{-0.25cm}
\section*{Acknowledgments}
% \vspace*{-0.25cm}
This work is supported by the 
U.S. Department of Energy under contract DE-FG02-87ER40365 %NTC  
and  
the National Science Foundation under contract PHY0070368.  %NTC
%.......................................................
%
%   REFERENCES
%
%.......................................................

\end{document}